\documentclass[preprint,aps,amsmath,superscriptaddress,nofootinbib,tightenlines]{revtex4}
\usepackage{graphicx}
\usepackage{bm}
\usepackage{epsfig}

\newif\ifpdf
\ifx\pdfoutput\undefined
\pdffalse 
\else
\pdfoutput=1 
\pdftrue
\fi

\def\bsigma{\mbox{\boldmath $\sigma$}}

\def\OMIT#1{}

\newcommand{\nn}{\nonumber} 

\newcommand{\bn}{{\bar n}}
\newcommand{\bea}{\begin{eqnarray}}
\newcommand{\eea}{\end{eqnarray}}

\newcommand{\bnP}{\bar {\cal P}}
\newcommand{\ppP}{{\cal P}_\perp}

\newcommand{\cP}{{\cal P}}

\newcommand{\mups}{M_\Upsilon}

\begin{document}

\ifpdf
\DeclareGraphicsExtensions{.pdf, .jpg}
\else
\DeclareGraphicsExtensions{.eps, .jpg,.ps}
\fi


\preprint{\vbox{ \hbox{CMU-HEP-02-13}   \hbox{FERMILAB-Pub-02/295-T} }}

\title{The Photon Spectrum in $\Upsilon$ Decays} 

\author{Sean Fleming}
\affiliation{Department of Physics, Carnegie Mellon University,
      	Pittsburgh, PA 15213\footnote{Electronic address: spf@andrew.cmu.edu}
	\vspace{0.1cm}}
	
\author{Adam K. Leibovich}
\affiliation{Theory Group, Fermilab, 
	P.O. Box 500, 
	Batavia, IL 
	60510\footnote{Electronic address: adam@fnal.gov}\vspace{0.2cm}}

\date{\today\\ \vspace{1cm} }



\begin{abstract}

We present  a theoretical prediction for the photon spectrum in 
radiative $\Upsilon$ decay. Parts of the spectrum have already been
understood, but an understanding of the endpoint region has 
remained elusive. In this paper we provide the missing piece, and 
resolve a controversy in the literature. We 
treat the endpoint region of $\Upsilon \to X \gamma$ decay within 
the framework of the soft-collinear effective theory (SCET). Within this 
approach the $\Upsilon$ structure function arises naturally, and kinematic
logarithms are summed by running operators using renormalization
group equations. In a previous paper we studied the color-octet
contribution to the decay. Here we treat the color-singlet
contribution. We combine our result with previous results to obtain
the $\Upsilon \to X \gamma$ spectrum. We find that resumming the 
color-singlet contribution in the endpoint gives a result  that is in much
better agreement with the data than the leading order  prediction.

\end{abstract}

\maketitle

\newpage
\section{Introduction}

The large mass of the constituent quarks in quarkonium makes this
system simple enough to use as a testing ground for theory.  For
example, the spectrum of quarkonium has been extensively studied in
Lattice QCD, allowing qualitative investigations of systematic errors
and extractions of parameters (such as the strong coupling constant
and the $b$ quark mass).  In addition, the gross under-estimate of the
production of $\psi'$  at the Tevatron helped establish the validity 
of a Non-Relativistic effective field theory of QCD (NRQCD)~\cite{bbl,lmr}.
While much has been learned about the charmonium and bottomonium
systems, it is still a useful probe for theory.

Inclusive decays of quarkonium are understood in the framework of the
operator product expansion (OPE), with power-counting rules given by 
NRQCD. The OPE for the direct photon spectrum of $\Upsilon$ decay 
is~\cite{bbl}
\begin{equation} \label{nrqcdope}
\frac{d \Gamma}{d z} = \sum_n C_n (M,z) 
   \langle \Upsilon \vert {\cal O}_n \vert \Upsilon \rangle \,,
\end{equation}
where $z = 2 E_\gamma /M$, with $M = 2 m_b$. The $C_i$ are
short-distance Wilson coefficients which can be calculated as a
perturbative series in $\alpha_s(M)$, and the ${\cal O}$ are NRQCD
operators. NRQCD power counting assigns a power of the relative
velocity, $v \ll 1$, of the heavy quarks to each operator, which
organizes the sum into a power series in $v$. At leading order in $v$
only one term in the sum must be kept, the so called color-singlet
contribution. The color-singlet operator ${\cal O}_1({}^3S_1)$ creates
and annihilates a quark-antiquark pair in a color-singlet ${}^3S_1$
configuration, and is multiplied by the color-singlet Wilson
coefficient, which at leading order is proportional to
$\alpha^2_s(M)$.

This simple picture of the photon spectrum in inclusive $\Upsilon$
decays is only valid in the intermediate range of the photon energy
spectrum ($0.3 \lesssim z \lesssim 0.7$). In the lower range,
$z\lesssim 0.3$, photon-fragmentation contributions are
important~\cite{Catani:1995iz, Maltoni:1999nh}. At large values of the
photon energy, $ z\gtrsim 0.7$, both the perturbative
expansion~\cite{Maltoni:1999nh} and the OPE~\cite{Rothstein:1997ac}
break down.

The breakdown of the OPE and the perturbative expansion is a
consequence of NRQCD not containing the correct low energy degrees of
freedom to describe the endpoint of the photon spectrum. The effective
theory which correctly describes this kinematic regime is a
combination of NRQCD for the heavy degrees of freedom, and the
soft-collinear effective theory
(SCET)~\cite{Bauer:2001ew,Bauer:2001yr,Bauer:2001ct,Bauer:2001yt} for
the light degrees of freedom. In a previous paper~\cite{Bauer:2001rh}
we applied SCET to the color-octet contributions to radiative
$\Upsilon$ decay. First we used the SCET power counting to show that
in the endpoint region there is a color-octet contribution which is
the same order as the color-singlet contribution. As discussed above
this is not what pure NRQCD power counting gives.  We then showed that
in SCET the octet structure functions arise naturally, and that
Sudakov logarithms are summed using the renormalization group
equations (RGEs). However, before a meaningful comparison to data can
be made the color-singlet contribution at the endpoint must also be
treated within SCET. This is the purpose of this work, which is an
expanded version of Ref.~\cite{Fleming:2002rv}.  

The logarithms at the endpoint for the color-singlet rate were previously 
studied by Photiadis~\cite{Photiadis:1985hn} and Hautmann~\cite{Hautmann:2001yz}.
We find that these logarithms come entirely from collinear physics, and 
can be summed into the form
\begin{equation}
{\rm exp} \bigg[ \sum_n a_n \alpha_s^n(M) \,  \ln^n (1-z) \bigg] \,.
\end{equation}
Our result agrees with that of Photiadis. 
Hautmann argues that all logarithms cancel in the color-singlet
contribution. We do not agree with that statement. However,
the analysis in Ref.~\cite{Hautmann:2001yz} was based on the
eikonal approximation, which is valid for soft physics. We do find 
that there are no logarithms arising from physics at this scale, only 
from the collinear scale.

In Section II we review the tree level spectrum, as calculated in
NRQCD, and the fragmentation results.  In Section III we match onto
SCET and present the leading operators consistent with the symmetries
of SCET. There are two color-octet operators and one color-singlet
operator. In Section IV we show how the color-singlet rate factorizes
into hard, jet and usoft functions. In Section V we perform an OPE
by integrating out collinear modes, and derive the tree-level, direct
rate in the $z\to1$ limit.  In Section VI we sum the Sudakov logarithms
in the color-singlet rate, by running the color-singlet operator using
the RGEs in SCET. By using the resummed coefficient for the
color-singlet operator from the previous section, we have a prediction
for the resummed rate in the endpoint region.  In Section VII we
discuss the phenomenology of the $\Upsilon$ radiative decay rate,
including the results of the previous sections.  We also compare our
results to calculations in the literature.  Finally we conclude in
Section VIII.

\section{Tree level spectrum and fragmentation}

Before we venture to the endpoint of the photon spectrum in inclusive
radiative $\Upsilon$ decay we review the theoretical results in the
lower and intermediate range of the photon energy spectrum, $z
\lesssim 0.7$. The leading order color-singlet contribution to the
decay rate was first calculated in Refs.~\cite{firstRad}. The
conventional wisdom at the time was that this process is computable
within QCD. However, Catani and Hautmann
\cite{Catani:1995iz} 
pointed out that there is a non-perturbative contribution which becomes
important at low $z$. This contribution is due to the hadronic content
of the photon, and makes itself noticed in perturbation theory through
the presence of infrared divergences. Catani and Hautmann showed that
these divergences can be absorbed into a non-perturbative photon
structure function $D(z,M)$.  A consequence of their analysis is that
the photon spectrum in $\Upsilon \to X \gamma$ can be written as a sum
of a direct contribution and a fragmentation contribution,
\begin{equation}
\frac{d\Gamma}{dz}= \frac{d\Gamma^{\rm dir}}{dz}
  +\frac{d\Gamma^{\rm frag}}{dz},
\end{equation}
where the direct term includes all contributions where the photon is
produced in the hard scattering, and the fragmentation term is the
contribution when the photon fragments from a parton produced in the
initial hard scattering.  We will look at each contribution in turn.

\subsection{The direct rate}

In NRQCD, the direct contribution can be calculated as an expansion in
$\alpha_s(M)$, where $M=2m_b$ is the $\Upsilon$ mass, and in $v$, the
relative velocity of the $b$ quarks inside the bound state.  The rate
is written as
\begin{equation}
\frac1{\Gamma_0}\frac{d\Gamma^{\rm dir}}{dz} = \sum_n C_n(M,z) 
   \langle \Upsilon \vert {\cal O}_n \vert \Upsilon \rangle,
\end{equation}
where the $C_n(z,M)$ are short distance Wilson coefficients,
calculable in perturbation theory, and the NRQCD matrix elements scale
with a certain power in $v$.  The lowest order contribution is the 
color-singlet $^3S_1$ operator. The matrix element of this operator 
can be related to the $\Upsilon$ wavefunction at the origin
\begin{eqnarray}\label{singletWF}
\langle \Upsilon \vert {\cal O}_1(^3S_1) \vert \Upsilon \rangle &=&
 \langle \Upsilon\vert\psi_{\bf p}^\dagger \bsigma_i\chi_{-\bf p}
\chi_{-{\bf p}'}^\dagger\bsigma_i\psi_{\bf p'}\vert\Upsilon\rangle
\nn\\
&=&  \frac{N_c}{2\pi} |R(0)|^2,
\end{eqnarray}
where $\psi^\dagger_{\bf p}$ and $\chi_{-\bf p}$ creates a heavy quark and antiquark,
respectively. 

The direct contribution was first calculated in the color-singlet
model in Ref.~\cite{firstRad}, which for $\Upsilon$ decay is
equivalent to the leading order in $v$ in NRQCD.  At lowest order in
$\alpha_s$, the rate is given by
\begin{equation}\label{LOrate}
\frac1{\Gamma_0} \frac{d\Gamma^{\rm dir}_{\rm LO}}{dz} =  
\frac{2-z}{z} + \frac{z(1-z)}{(2-z)^2} +
2\frac{1-z}{z^2}\ln(1-z) - 2\frac{(1-z)^2}{(2-z)^3} \ln(1-z),
\end{equation}
where 
\begin{equation}
\Gamma_0 = \frac{32}{27}\alpha\alpha_s^2e_b^2
\frac{\langle\Upsilon\vert{\cal O}_1(^3S_1)\vert\Upsilon\rangle}{m_b^2},
\label{gamma0}
\end{equation}
and $e_b = -1/3$.  The $\alpha_s$ correction to this rate was
calculated numerically in Ref.~\cite{Kramer:1999bf}. It leads to small
corrections over most of phase space; however in the endpoint region
the corrections are of order the leading order (LO) contribution.

At higher order in the velocity expansion, there are direct
contributions from the color-octet matrix elements
\cite{Maltoni:1999nh}.  The decay through a color-octet matrix element
can occur at one lower order in $\alpha_s$, with the $b\bar{b}$
decaying to a photon and gluon.  At order $v^4$ and lowest order
in $\alpha_s$, the decay rates are
\begin{eqnarray}
\frac{d\Gamma^{\rm dir}(^1S_0^{(8)})}{dz} &=& 4\pi e_b^2\alpha\alpha_s
\frac{\langle\Upsilon\vert{\cal O}_8(^1S_0)\vert\Upsilon\rangle}{m_b^2}
\delta(1-z),\\
\frac{d\Gamma^{\rm dir}(^3P_J^{(8)})}{dz} &=& 28\pi e_b^2\alpha\alpha_s
\frac{\langle\Upsilon\vert{\cal O}_8(^3P_0)\vert\Upsilon\rangle}{m_b^4}
\delta(1-z),
\label{oct3pj}
\end{eqnarray}
where we have summed over $J=0,2$ in Eq.~(\ref{oct3pj}) and used the
relation 
\begin{equation}
\langle\Upsilon\vert{\cal O}_8(^3P_J)\vert\Upsilon\rangle = (2J+1)
\langle\Upsilon\vert{\cal O}_8(^3P_0)\vert\Upsilon\rangle.
\end{equation}
At order $\alpha_s$, these are the only contributions.  Note that
these contributions are singular at the upper endpoint.  The 
next-to-leading order (NLO)
$\alpha_s^2$ color-octet contributions were also calculated in
\cite{Maltoni:1999nh}.  At this order, there are also contributions from
the color-octet $^3S_1$ and $^3P_1$ channels.  Large Sudakov
logarithms appear in the $^1S_0$ and $^3P_{0,2}$ contributions near
the upper endpoint, and have been resummed in \cite{Bauer:2001rh}.
We will discuss this resummation below.

\subsection{Fragmentation contribution}

Catani and Hautmann pointed out the importance of fragmentation for
the photon spectrum in quarkonium decays \cite{Catani:1995iz}.  The
fragmentation rate can be written as
\begin{equation}
\frac{d\Gamma^{\rm frag}}{dz} = 
  \sum_{a = q,\bar q, g} \int_z^1\frac{dx}{x} \frac{d\Gamma_a}{dx}
    D_{a\gamma}\left(\frac{z}{x},M\right),
\end{equation}
where the rate to produce parton $a$, $d\Gamma_a/dx$, is convoluted
with the probability that the parton fragments to a photon,
$D_{a\gamma}$, with energy fraction $z/x$.  The rate to produce parton
$a$ can again be expanded in powers of $v$ \cite{Maltoni:1999nh}, with
the leading term being the color-singlet rate for an $\Upsilon$ to
decay to three gluons,
\begin{equation}
\frac{d\Gamma^{\rm frag}_{\rm LO}}{dz} = 
\int_z^1\frac{dx}{x} \frac{d\Gamma_{ggg}}{dx} 
D_{g\gamma}\left(\frac{z}{x},M\right).
\label{CSfrag}
\end{equation}
The rate to three gluons can be obtained from Eq.~(\ref{LOrate}) by a
change of color-factors,
\begin{equation}
\frac{d\Gamma_{ggg}}{dz} = 
  \frac{5}{12}\frac{\alpha_s}{e_b^2\alpha} 
  \frac{d\Gamma^{\rm dir}_{\rm LO}}{dz}.
\end{equation}
At order $v^4$ there are three color-octet fragmentation
contributions, with the order $\alpha_s^2$ partonic rates 
being~\cite{Maltoni:1999nh}
\begin{eqnarray}
\frac{d\Gamma(^1S_0^{(8)}\to gg)}{dz} &=&
\frac{5\pi\alpha_s^2}{3}
\frac{\langle\Upsilon\vert{\cal O}_8(^1S_0)\vert\Upsilon\rangle}{m_b^2}
\delta(1-z),\\
\frac{d\Gamma(^3P_J^{(8)}\to gg)}{dz} &=&
\frac{35\pi\alpha_s^2}{3}
\frac{\langle\Upsilon\vert{\cal O}_8(^3P_0)\vert\Upsilon\rangle}{m_b^4}
\delta(1-z),\\
\frac{d\Gamma(^3S_1^{(8)}\to q\bar q)}{dz} &=&
\frac{\pi\alpha_s^2}{3}
\frac{\langle\Upsilon\vert{\cal O}_8(^3S_1)\vert\Upsilon\rangle}{m_b^2}
\delta(1-z).
\end{eqnarray}
The $\alpha_s$ corrections to these have been calculated, and can be
found in Ref.~\cite{Petrelli:1997ge}.

These rates must be convoluted with the fragmentation functions,
$D_{a\gamma}(z,M)$.  The $M$-dependence of the fragmentation functions
can be predicted using perturbative QCD via Altarelli-Parisi evolution
equations.  However, the solution depends on non-perturbative
fragmentation function at some input scale $\Lambda$, which must be
measured from experiment.  This has been done by the ALEPH
collaboration for the $D_{q\gamma}$ fragmentation function
\cite{Buskulic:1995au}, which is parameterized as
\begin{equation}
D_{q\gamma}(z,\mu) = \frac{e_q^2\alpha(\mu)}{2\pi}
\left[P_{q\gamma}(z) \ln\left(\frac{\mu^2}{\mu_0^2(1-z)^2}\right) +
C\right],
\label{quarkfrag}
\end{equation}
where $C = -1-\ln(M_Z^2/(2\mu_0^2))$ and $P_{q\gamma}(z)$ is the
Altarelli-Parisi splitting function \cite{Altarelli:1977zs},
\begin{equation}
P_{q\gamma}(z) = \frac{1+ (1-z)^2}{z}.
\end{equation}
The value for $\mu_0$ extracted from the data is
\begin{equation}
\mu_0 = 0.14^{+0.43}_{-0.12} {\rm\ GeV}.
\end{equation}
The gluon to photon fragmentation function has not yet been measured.
Here we show the parameterization due to Owens \cite{Owens:1986mp},
which uses the approximation $D_{a\gamma}(z,\Lambda_{\rm QCD}) = 0$,
\begin{eqnarray}
z D_{g\gamma}(z,M) &=& \frac{\alpha}{2\pi} 0.0243 (1-z)
z^{-0.97}\ln(M^2/\Lambda_{QCD}) \nonumber\\
&=&\frac{2\alpha}{\alpha_s(M)\beta_0} 0.0243 (1-z)z^{-0.97},
\end{eqnarray}
where $\beta_0 = 11-2n_f/3$ is the QCD beta function.

\section{Matching onto  SCET}\label{matchsection}

Next we turn our attention to the endpoint region. The NRQCD
power-counting rules break down in this regime because NRQCD does not
include the appropriate long distance modes: collinear physics is
missing from the theory. An effective theory which does include
collinear physics is 
SCET~\cite{Bauer:2001ew,Bauer:2001yr, Bauer:2001ct,Bauer:2001yt}. 
This theory describes the interactions of highly energetic collinear modes
with soft degrees of freedom. To describe $\Upsilon$ decay at the
endpoint we have to couple SCET with NRQCD.

Before we describe how to match onto SCET it is helpful to understand
the scales which arise in the problem. Consider the momentum of a
collinear particle moving near the lightcone. In lightcone coordinates
we can write this momentum as $p=(p^+,p^-,p_\perp)$. Since the mass of
the particle is much smaller than its energy, we define $p^2 \sim M^2
\lambda^2$, where $M$ is the scale that sets the energy and $\lambda$
is a small parameter. The lightcone momentum components of the
collinear particle are widely separated. If we choose $p^-$ to be
${\cal O}(M)$, then $p_\perp/p^- \sim \lambda$, and $p^+/p^- \sim
\lambda^2$. We refer to the latter two scales as collinear and
ultrasoft (usoft), respectively. To be concrete consider the $b \bar
b$ pair to have momentum $Mv^\mu+k^\mu$, where $v^\mu = (1,0,0,0)$ and
$k^\mu$ is ${\cal O}(\Lambda_{\rm QCD})$ in the $\Upsilon$
center-of-mass frame. The photon momentum is $Mz\bar{n}^\mu /2$, where
we have chosen $\bar{n}^\mu = (1,0,0,1)$. In the endpoint region the
hadronic jet recoiling against the photon moves in the opposite
lightcone direction $n^\mu = (1,0,0,-1)$, with momentum $p^\mu_X =
Mn^\mu /2 + M(1-z) \bar{n}^\mu /2 + k^\mu$ . Thus the hadronic jet has
$\bar{n}\cdot p_X = p^-_X \sim M$. Next note that $m^2_X \approx
M^2(1-z)$. For $(1-z) \sim v^2 \sim \Lambda_{\rm QCD}/M$ we find
\begin{eqnarray}
m_X \sim \sqrt{M \Lambda_{\rm QCD}}\,,
\end{eqnarray}
which is the collinear scale. This implies that for this process
the collinear-soft expansion parameter $\lambda$ is of order
$\sqrt{1-z}\sim\sqrt{\Lambda_{\rm QCD}/M}$. 
The usoft scale is the component of the
hadronic momentum in the $n$ direction:
\begin{eqnarray}
n \cdot p_X \sim \frac{m^2_X}{\bar{n} \cdot p_X} 
\sim \Lambda_{\rm QCD} \sim M \lambda^2\,.
\end{eqnarray}

By matching onto SCET the large scale $M$ is integrated out.  In
practice, the matching procedure is to calculate matrix elements in
QCD, expand them in powers of $\lambda$, and match onto products of
Wilson coefficients and operators in SCET. Thus it is important to be
able to deduce the SCET operators which can arise at a given order in
$\lambda$.  Field theory generally allows any operators that are
consistent with the symmetries of the theory.  As explained in detail
in Ref.~\cite{Bauer:2001yt}, the symmetry of SCET which restricts the
operators that can arise is the combined collinear- and usoft-gauge
invariance of the theory.

We will use the usoft- and collinear-gauge invariance of SCET to obtain
the operators which are needed for the endpoint distribution of
$\Upsilon \to X \gamma$ decays.  However, we first review the building
blocks from which these operators are constructed. In SCET there are 
fundamental fields and Wilson lines, which are built out of the fields. 
Furthermore there are two separate sectors to the theory: collinear and
usoft.\footnote{Soft modes do not enter  our analysis so we do not
include them.}  
In the collinear sector there is a collinear fermion field
$\xi_{n,p}$, a collinear gluon field $A_{n,q}^\mu$, and a collinear
Wilson line
\begin{equation}
W_n(x)=
 \bigg[ \sum_{\rm perms} {\rm exp} 
  \left( -g_s \frac{1}{\bnP} \bn \cdot A_{n,q}(x) \right) \bigg] \,.
\end{equation}
The subscripts on the collinear fields are the lightcone direction
$n^\mu$, and the large components of the lightcone momentum ($\bn\cdot
q, q_\perp$). The operator ${\cal P}^\mu$ projects out the momentum
label~\cite{Bauer:2001ct}.  For example $\bn\cdot {\cal P} \xi_{n,p}
\equiv \bnP \xi_{n,p} = \bn \cdot p \xi_{n,p}$.  Likewise in the usoft
sector there is a usoft fermion field $q_s$, a usoft gluon field
$A^\mu_s$, and a usoft Wilson line $Y$. Operators in SCET are
constructed out of these objects such that they are gauge invariant.
For example under collinear-gauge transformations $\xi_{n,p} \to U_n
\xi_{n,p}$, and $W_n \to U_n W_n$, so the combination
\begin{equation}
\chi_n \equiv W^\dagger_n \xi_{n,p}
\end{equation}
is collinear-gauge invariant. This combination, however, still
transforms under a usoft-gauge transformation $\chi_n 
\to V(x) \chi_n $. A collinear-gauge invariant field
strength is
\begin{equation}
G^{\mu\nu}_n \equiv -\frac{i}{g_s} W^\dagger [i{\cal D}_n^\mu 
  + g_sA_{n,q}^\mu, i{\cal D}_n^\nu+g_sA_{n,q'}^\nu ] W \,,
\end{equation}
where 
\begin{equation}
i{\cal D}_n^\mu = \frac{n^\mu}2 \bnP + \ppP^\mu + 
\frac{\bn^\mu}2 i n\cdot D,
\end{equation}
and $iD^\mu = i \partial^\mu+g_sA^\mu_s$ is the usoft covariant
derivative.  Note that $G^{\mu\nu}_n$ is not homogeneous in the power
counting. The leading piece scales like $\lambda$, and is given by
$\bnP B^\mu_\perp\equiv\bn_\nu G^{\nu\mu}_n$, where the perp subscript
on $B$ indicates that the $\mu$ index only has support over
perpendicular components. Simplifying we obtain
\begin{equation}\label{bfield}
B^\mu_\perp =  \frac{-i}{g_s} W^\dagger (\ppP^\mu + g_s (A^\mu_{n,q})_\perp)W. 
\end{equation}
We use these objects to build the operators we need to match onto SCET
at the endpoint of the $\Upsilon\to X\gamma$ spectrum. For further
examples the reader is referred to Ref.~\cite{Bauer:2002nz}.

We are now in a position where we can write down the leading
operators.  Aside from $B_\perp$, we will also need the heavy quark
and antiquark fields, $\psi_{\bf p}(x)$ and $\chi_{-{\bf p}}(x)$, from
NRQCD.  Consider first the color-octet ${}^1S_0$
operator~\cite{Bauer:2001rh}. The heavy quark-antiquark pair must be
in an octet ${}^1S_0$ state, and we know that at the endpoint this
pair annihilates into a photon and a jet. Since the operator with a 
quark jet has a vanishing tree level Wilson coefficient we do not consider
it here. We only consider the gluon jet operator. So our operator must
include the collinear-gauge invariant field strength. We want the
operator to be homogeneous in the power counting, so it is
$B^\mu_\perp $ that we use to build our operator.  The most general LO
color-octet ${}^1S_0$ operator which is consistent with these
requirements, and is usoft- and collinear-gauge invariant is
\begin{equation} \label{1s0op1}
   \chi^\dagger_{- {\bf p}} \, \Gamma^{(8,{}^1S_0)}_{\alpha \mu}(-\bnP,\mu)  
 B^\alpha_\perp \psi_{\bf p} \,,
\end{equation}
where $\Gamma^{(8,{}^1S_0)}_{\alpha \mu}(-\bnP)$ is the matching
coefficient. It is a function of the renormalization scale $\mu$, and
the large lightcone component of the collinear momentum projected out
by $\bnP$. By momentum conservation the large lightcone momentum
component must be equal to $-M$, so $\bnP B^\alpha_\perp = -M
B^\alpha_\perp$, and Eq.~(\ref{1s0op1}) simplifies to
\begin{equation}\label{1s0op2}
\Gamma^{(8,{}^1S_0)}_{\alpha \mu}(M,\mu) \,
 \chi^\dagger_{-{\bf p}} B^\alpha_\perp \psi_{\bf p} \,.
\end{equation}
Note this operator is order $\lambda$, and there is no operator of
lower order in $\lambda$ which can be constructed.

\begin{figure}[t]
\centerline{\includegraphics[width =4in]{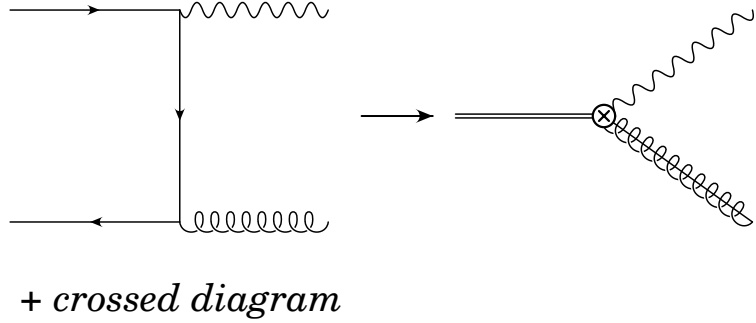}}
\caption{\it  Matching the decay amplitude for $b\bar{b} \to \gamma +
g$ in QCD and SCET.  Collinear gluons are represented by a spring with
a line through it.  
\label{1s0matchingfig}}
\end{figure}

We determine the matching coefficient by requiring the SCET amplitude
for a color-octet ${}^1S_0$ $b\bar{b}$ pair decaying into a photon and
collinear gluons to be equal to the QCD amplitude expanded to order
$\lambda$. This matching is carried out order by order in an expansion
in $\alpha_s(M)$.  As an explicit example we determine
$\Gamma^{(8,{}^1S_0)}_{\alpha \mu}(M)$ to leading order in
$\alpha_s$. The Feynman graphs are shown in Fig.~\ref{1s0matchingfig}.
The amplitude for a color-octet ${}^1S_0$ $b\bar{b}$ pair to decay
into a photon and a collinear gluon is simply given by the Feynman
rule for the color-octet ${}^1S_0$ operator in Eq.~(\ref{1s0op2})
\begin{equation}\label{1s0feynman} 
\Gamma^{(8,{}^1S_0)}_{\alpha \mu}(M) 
 \Big( \epsilon^\alpha_\perp 
 - \frac{q^\alpha_\perp}{\bn\cdot q} \bn\cdot\epsilon \Big) 
\eta^\dagger_{-{\bf p}}  T^A  \xi_{\bf p} \,,
\end{equation}
where $\eta$ and $\xi$ are two component spinors, and momentum 
conservation sets $\bn\cdot q = M$.  The tree level QCD
expression for this amplitude expanded to order $\lambda$ is
\begin{equation}\label{1s0treeQCD}
2 g_s e e_b \epsilon_{\alpha \mu}^\perp  
\Big( \epsilon_\perp^\alpha 
  - \frac{q_\perp^\alpha}{\bn\cdot q} \bn\cdot\epsilon \Big)
\eta^\dagger_{-{\bf p}} T^A \xi_{\bf p} \,,
\end{equation}
where $\epsilon^{\alpha\mu}_\perp
=\epsilon^{\alpha\mu\rho\beta}\bar{n}_\rho v_\beta$.  Note there is no
$\lambda^0$ piece.  Setting the two expressions equal to each other we
obtain
\begin{equation}\label{1s0matching}
 \Gamma^{(8,{}^1S_0)}_{\alpha \mu}(M)  
   = 2 g_s e e_b \epsilon_{\alpha \mu}^\perp  \,.
\end{equation}
The color-octet ${}^3P_J$ operator is also order $\lambda$.  We give
that operator and the matching coefficient in Appendix~\ref{appFR}.  At
order $\lambda$ these are the only two possible operators that can be
constructed.  

\begin{figure}[t]
\centerline{\includegraphics[width=4in]{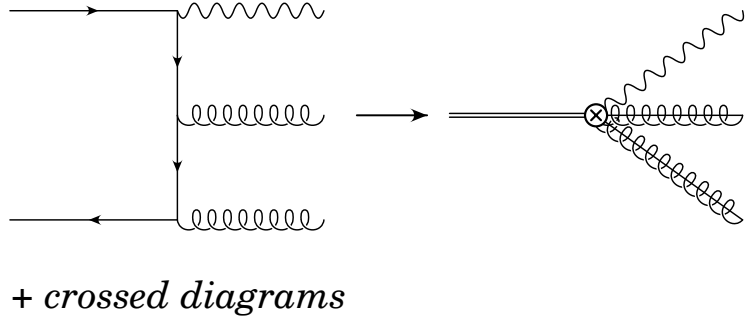}}
\caption{\it Matching the decay amplitude for $b\bar{b} \to \gamma +
gg$ in QCD and SCET.
\label{3s1matchingfig}}
\end{figure}

Next we construct the color-singlet ${}^3S_1$ operator at the
endpoint. A $b\bar{b}$ pair in a color-singlet ${}^3S_1$ configuration
decays into a photon and a collinear jet of gluons that must be
colorless.\footnote{We could also have an operator where the
$b\bar{b}$ pair decays to a photon and a quark-antiquark pair.  The
coefficient for this operator is zero at tree level, but would mix
with the operator we have here.  Numerically the effect of this mixing
is small, and we will look into its effect
in~\cite{FL}.\label{noquarkfoot}} The only way to construct such an
operator is to include two of the $B_\perp$ fields in a colorless
configuration. The only operator that can be constructed such that it
is collinear- and usoft-gauge invariant is
\begin{equation}\label{3s1op1}
{\cal O}(1,{}^3S_1) =
 \chi^\dagger_{-{\bf p}} \Lambda\cdot\bsigma^\delta \psi_{\bf p}
{\rm Tr} \big\{ B^\alpha_\perp \, 
\Gamma^{(1,{}^3S_1)}_{\alpha \beta \delta \mu} ( \bnP, \bnP^\dagger ) \, 
B^\beta_\perp \big\} \,,
\end{equation}
where $\bnP^\dagger$ operates on the fields to the left, and $\Lambda$
boosts from the $\Upsilon$ rest frame to a frame where the $\Upsilon$
has arbitrary four-momentum~\cite{bc}.  This operator is ${\cal
O}(\lambda^2)$ in the power counting. Once again momentum conservation
forces the total momentum of the jet to be $M$, so that
$B^\alpha_\perp ( \bnP+ \bnP^\dagger) B^\beta_\perp = -M B^\alpha_\perp
B^\beta_\perp$.  Introducing $\cP_{-} = \bnP- \bnP^\dagger$,
Eq.~(\ref{3s1op1}) can be simplified to
\begin{equation}\label{3s1op2}
{\cal O}(1,{}^3S_1) =
\chi^\dagger_{-{\bf p}} \Lambda\cdot\bsigma^\delta \psi_{\bf p}
{\rm Tr} \big\{ B^\alpha_\perp \, 
\Gamma^{(1,{}^3S_1)}_{\alpha \beta \delta \mu} ( M,\cP_{-} ) \, 
B^\beta_\perp \big\} \,.
\end{equation}
The tree matching is shown in Fig.~\ref{3s1matchingfig}.  
We obtain
\begin{eqnarray}\label{matching1}
\Gamma^{(1,{}^3S_1)}_{\alpha \beta \delta \mu} ( M,\bn\cdot q_{-})
&=& \frac{4 g_s^2 e e_b}{3 M} g_{\alpha\beta}^\perp \Big[ g_{\mu\delta}+
 \frac{1}{2} n_\delta n_\mu  \Big]
\nn \\
&=& \frac{4 g_s^2 e e_b}{3 M} g_{\alpha\beta}^\perp g_{\mu\delta} \,,
\end{eqnarray}
where $g_\perp^{\mu\nu} = g^{\mu\nu} - (n^\mu\bar n^\nu + n^\nu\bar
n^\mu)/2$, $\bn\cdot q_{-} = \bn\cdot q- \bn\cdot q'$, and the last
line holds if the photon is transverse.

\section{Factorization in Inclusive $\Upsilon \to X \gamma$}

Next we show that in the endpoint region the inclusive $\Upsilon \to X
\gamma$ decay rate can be factored into a hard coefficient, a
collinear jet function, and a usoft function.  We then perform an OPE
by integrating out collinear modes at the scale $\mu_c \approx M
\sqrt{1-z} \gg \Lambda_{\rm QCD}$, and match onto a usoft operator
convoluted with a matching function. The usoft operator is nonlocal
along one lightcone direction, and, as pointed out by Rothstein and
Wise in Ref.~\cite{Rothstein:1997ac}, can be calculated for the
color-singlet contribution.

The factorization proof for $\Upsilon \to X \gamma$ is similar to that of
$b \to X_s \gamma$, and we follow Refs.~\cite{Bauer:2001yt,Bauer:2002nz}.
 Using the optical theorem the
inclusive photon energy spectrum can be written as
\begin{equation}\label{optcthm}
\frac{d \Gamma}{d z} = z \frac{M}{16 \pi^2} {\rm Im} T(z)  \,,
\end{equation}
where the forward scattering amplitude $T(z)$ is 
\begin{equation}\label{fsamp}
 T(z) = -i \int d^4 x e^{-iq\cdot x} 
\langle \Upsilon | T J^\dagger_\mu (x) J_\nu(0) | \Upsilon \rangle 
   g^{\mu \nu}_\perp \,.
\end{equation}
The $\Upsilon$ states are relativistically normalized, and the $T$
indicates time ordering. We will show that in the endpoint region, at
leading order in the SCET power counting the decay rate can be
expressed in a factored form to all orders in $\alpha_s$
\begin{equation}\label{factthmI}
\frac{d \Gamma}{d z}  = 
\sum_\omega H(M,\omega, \mu) \int dk^+ S(k^+,\mu) \;
  {\rm Im}\, J_\omega(k^+ +M(1-z),\mu) \,.
\end{equation}

The first step is to match the QCD current $J_\mu$ in Eq.~(\ref{fsamp})
onto SCET operators. The leading order in $\lambda$ operators are the
color-octet ${}^1S_0$ and ${}^3P_J$ operators. At one order higher in
$\lambda$ we match onto the color-singlet operator,
Eq.~(\ref{3s1op1}). We will show in this section how the color-singlet
contribution factorizes, and leave the color-octet contribution for
Appendix~\ref{facappen}.

Matching the current to leading order in $\lambda$ we obtain
\begin{equation}\label{currmatch2}
J_\mu = \sum_{\omega}   e^{-i(Mv+\bnP \frac{n}{2} )\cdot x}
i \Gamma^{(1,{}^3S_1)}_{\alpha \beta \delta \mu}(\omega) \,  
\tilde{J}^{\delta \alpha \beta}_{(1,{}^3S_1)}(\omega)  \,,
\end{equation}
where the effective current is
\begin{equation}
\tilde{J}^{\delta \alpha \beta}_{(1,{}^3S_1)}(\omega) = 
\chi^\dagger_{-{\bf p}} \Lambda\cdot\bsigma^\delta \psi_{\bf p}
{\rm Tr} \big\{ B^\alpha_\perp \, 
  \delta_{ \omega,\cP_{-}} \, B^\beta_\perp \big\},
\end{equation}
and $\Gamma^{(1,{}^3S_1)}_{\alpha \beta \delta \mu}(\omega)$ is given
in Eq.~(\ref{matching1}).
Note momentum conservation forces $\bnP \tilde{J}(\omega) = -M
\tilde{J}(\omega)$, so we can replace $\bnP$ by $-M$ in the phase
factor.  Substituting Eq.~(\ref{currmatch2}) into Eq.~(\ref{fsamp})
gives
\begin{equation}\label{csbot}
T(z) =  
\sum_{\omega,\omega'} 
H_{\delta \delta' \alpha \alpha' \beta \beta'}(\omega,\omega')
T^{\delta\delta'\alpha\alpha'\beta\beta'}_{\rm eff}(\omega,\omega', z,\mu) \,,
\end{equation}
where
\begin{equation}
T^{\delta\delta'\alpha\alpha'\beta\beta'}_{\rm eff}(\omega,\omega', z, \mu)  = 
-i \int d^4 x e^{i(M\frac{\bn}{2} - q)\cdot x} 
\langle \Upsilon | 
  T \tilde{J}^{\delta \alpha \beta \dagger}_{(1,{}^3S_1)}(\omega,x)
  \tilde{J}^{\delta' \alpha' \beta'}_{(1,{}^3S_1)}(\omega',0) |  
\Upsilon \rangle \,,
\end{equation}
and to leading order in $\alpha_s(M)$ we obtain
\begin{equation}
H_{\delta \delta' \alpha \alpha' \beta \beta'}(\omega,\omega')
= -\left(\frac{4 g_s^2 e e_b}{3M}\right)^2 
g^\perp_{\alpha\beta}g^\perp_{\alpha'\beta'}g^\perp_{\delta\delta'}.
\end{equation}

The usoft gluons in $T^{\rm eff}$ can be decoupled from the collinear
fields as explained in Ref.~\cite{Bauer:2001yt} by making the
field redefinition
\begin{equation}\label{fieldred}
A^\mu_{n,q} = Y A^{(0) \mu}_{n,q} Y^\dagger
\hspace{.5cm} \to \hspace{.5cm} 
W = Y W^{(0)} Y^\dagger \,,
\end{equation}
where the first identity implies the second.
The fields with the superscript $(0)$ do not interact with usoft
physics.  In the color-singlet contribution all usoft Wilson lines $Y$
cancel due to the identity $Y^\dagger Y = 1$. Furthermore  the
$\Upsilon$ state contains no collinear quanta. 
Thus we can separate the collinear physics
from the usoft physics and write the effective forward scattering
amplitude as
\begin{eqnarray}
T^{\delta\delta'\alpha\alpha'\beta\beta'}_{\rm eff}(\omega,\omega', z, \mu) 
& = & -i \int d^4 x e^{\frac{i}{2}M(1-z)\bn\cdot x}
\langle \Upsilon | 
 T \big[\psi^{\dagger}_{\bf p}
        \Lambda\cdot\bsigma^\delta\chi_{-{\bf p}} \big] (x)
   \big[ \chi^{\dagger}_{-{\bf p}'} \Lambda\cdot\bsigma^{\delta'} 
      \psi_{{\bf p}'} \big] (0)  |  \Upsilon \rangle
\nn \\
&  &  \times
\langle 0 | T \, {\rm Tr}\big[ B^{(0) \alpha}_\perp 
  \delta_{\omega,\cP_-} B^{(0) \beta}_\perp \big] (x)
  {\rm Tr}\big[ B^{(0) \alpha'}_\perp \delta_{\omega',\cP_-} 
  B^{(0) \beta'}_\perp \big] (0) |0 \rangle \,,
\end{eqnarray}
where we have chosen a frame such that $q^\mu = \frac{M}{2} z \bn^\mu$.
The collinear and usoft matrix elements appearing on the left hand side
of the above equation can be simplified. Using rotational
invariance~\cite{bc}, the usoft matrix element can be written as
\begin{eqnarray}\label{cssoft1}
& &  \Lambda^\delta_i \Lambda^{\delta'}_j
\langle \Upsilon | 
  T \big[ \psi^{\dagger}_{\bf p} \bsigma^i \chi_{-{\bf p}} \big] (x)
\big[ \chi^{\dagger}_{-{\bf p}'} \bsigma^j \psi_{{\bf p}'} \big] (0) |  
\Upsilon \rangle =
\nn \\
& & \qquad \frac{1}{3} \delta^{ij} \Lambda^\delta_i \Lambda^{\delta'}_j 
\langle \Upsilon | 
  T \big[ \psi^{\dagger}_{\bf p} \bsigma^k \chi_{-{\bf p}} \big] (x)
  \big[ \chi^{\dagger}_{-{\bf p}'} \bsigma^k \psi_{{\bf p}'} \big] (0) |
\Upsilon \rangle \,.
\end{eqnarray}
We can then use the identity $\delta^{ij} \Lambda^\delta_i
\Lambda^{\delta'}_j = (v^\delta v^{\delta'} - g^{\delta \delta'})$,
where $v^\delta$ is the four-velocity of the $\Upsilon$. 

Next we define a color-singlet jet function
\begin{eqnarray}\label{cscoll1}
\langle 0 | T \, {\rm Tr}\big[ B^{(0) \alpha}_\perp 
  \delta_{\omega,\cP_-} B^{(0) \beta}_\perp \big] (x)
{\rm Tr}\big[ B^{(0) \alpha'}_\perp \delta_{\omega',\cP_-} 
  B^{(0) \beta'}_\perp \big] (0) |0 \rangle
& \equiv &
\nn \\
& & \hspace{-30ex} \frac{i}{2} 
(g^{\alpha \alpha'}_\perp g^{\beta \beta'}_\perp+ 
  g^{\alpha \beta'}_\perp g^{\beta \alpha'}_\perp)
\, \delta_{\omega, \omega'} \, \int \frac{d^4 k}{(2 \pi)^4} 
e^{-i k\cdot x} J_\omega(k^+,\mu) \,.
\end{eqnarray}
The jet function, $J_\omega(k^+,\mu)$, is only a function of one component
of the usoft momentum, $k^+$, which follows from the collinear
Lagrangian containing only the $n\cdot \partial$
derivative~\cite{Bauer:2001yt}.  The hard coefficient in
Eq.~(\ref{csbot}) will contract with the factors in
Eqs.~(\ref{cssoft1})~and~(\ref{cscoll1}) so that the expression for
the forward scattering amplitude can be written as
\begin{equation}\label{morebots}
T(z) = \sum_\omega H(\omega,\mu) T_{\rm eff} (\omega,z,\mu) \,,
\end{equation}
where
\begin{eqnarray}\label{csalmost}
T_{\rm eff} (\omega,z,\mu) &= & 
\int \frac{d^4 x}{2 \pi} \int d k^+ \delta(x^+) \delta^{(2)}(x_\perp)
e^{\frac{i}{2}[M(1-z)-k^+]x^-} J_\omega (k^+,\mu)
\nn \\
& & \qquad \times \langle \Upsilon | 
  T \big[\psi^\dagger_{\bf p} \bsigma_i \chi_{-{\bf p}} \big] (x)
\big[\chi^\dagger_{-{\bf p}'} \bsigma_i \psi_{{\bf p}'} \big] (0) | 
\Upsilon \rangle \,,
\end{eqnarray}
and the leading order hard function is
\begin{equation}\label{lohardfun}
H(\omega,\mu) = \frac{4}{3}\left(\frac{4g_s^2 e e_b}{3M}\right)^2\,.
\end{equation}
Finally we define a color-singlet usoft function
\begin{equation}\label{cssoftfn}
S(\ell^+,\mu) = 
\int \frac{d x^-}{4 \pi} e^{\frac{-i}{2} \ell^+ x^-}
 \langle \Upsilon | 
T \big[\psi^\dagger_{\bf p} \bsigma_i \chi_{-{\bf p}} \big] (x^-)
\big[\chi^\dagger_{-{\bf p}'} \bsigma_i \psi_{{\bf p}'} \big] (0) | 
\Upsilon \rangle \,,
\end{equation}
which when substituted into Eq.~(\ref{csalmost}) gives the desired
factored form
\begin{equation}
T_{\rm eff} (\omega,z,\mu)  = 
\int  d \ell^+  J_\omega[\ell^+ + M(1-z),\mu] S(\ell^+,\mu) \,.
\end{equation}
This proves the factorization theorem Eq.~(\ref{factthmI}).

\section{The Operator Product Expansion}

As we discussed in Section III the collinear scale in 
$\Upsilon \to X \gamma$ decay at the endpoint is of order 
$\sqrt{M \Lambda_{\rm QCD}} \sim 3 \, {\rm GeV}$. It is our opinion 
that this is large enough so that we can perform an OPE and integrate out
collinear physics. However one does not have to carry out this step. Then only 
hard contributions are integrated out, and both the jet function and the usoft 
function in Eq.~(\ref{factthmI}) are non-perturbative.  

The OPE is carried out by expanding the jet function in powers of 
$\alpha_s(M \lambda )$, and matching onto a non-local usoft 
operator of the form given in Eq.~(\ref{cssoftfn}) convoluted with a Wilson
coefficient.  The forward scattering amplitude will then be of the
form
\begin{equation}\label{nonotmore}
T(z) = \int d\ell^+ S(\ell^+,\mu) {\cal H}_J[\ell^+ + M(1-z),\mu] \,,
\end{equation}
where the Wilson coefficient is labeled with a $J$ to remind us that
it contains the hard part $H$ which is determined by matching SCET and
QCD in an expansion in $\alpha_s(M)$. In addition it contains the
perturbative expansion of the jet function in powers of
$\alpha_s(M\sqrt{1-z})$.  We now perform the OPE for the color-singlet
contribution, and finally use the results of Rothstein and
Wise~\cite{Rothstein:1997ac} to calculate the usoft function.

\begin{figure}[t]
\centerline{ \includegraphics[width=2.5in]{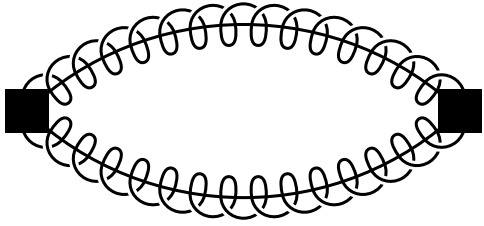}}
\caption{\it Feynman diagram for the leading order jet function.
\label{vacloop}}
\end{figure}

First we calculate the jet function, Eq.~(\ref{cscoll1}), to leading
order. The Feynman diagram for the vacuum matrix element is shown in
Fig.~\ref{vacloop}.
Evaluating the diagram gives 
\begin{eqnarray}\label{diagram3}
\langle 0 | T \, {\rm Tr}\big[ B^{(0) \alpha}_\perp 
\delta_{\omega,\cP_-} B^{(0) \beta}_\perp \big] (x)
{\rm Tr}\big[ B^{(0) \alpha'}_\perp \delta_{\omega',\cP_-} 
B^{(0) \beta'}_\perp \big] (0) |0 \rangle
& = & -2 
(g^{\alpha \alpha'}_\perp g^{\beta \beta'}_\perp+
g^{\alpha \beta'}_\perp g^{\beta \alpha'}_\perp) 
 \\
& & \hspace{-55ex}
\times \int \frac{d^4 K}{(2 \pi)^4}  
\int \frac{d^4 k}{(2 \pi)^4} 
\frac{
\delta_{\omega, \bn\cdot q}\, \delta_{\omega, \omega'} \,
e^{-iK\cdot x}}{[(M+\bn\cdot q)(K^+ + k^+) + q^2_{\perp}+i \delta][
(M-\bn\cdot q)(K^+ - k^+) + q^2_{\perp}+i \delta]} \,,
\nn
\end{eqnarray}
where we have used momentum conservation to set $\bn\cdot q_1 +
\bn\cdot q_2 = M$, and $ q^\mu_{1\perp} + q^\mu_{2\perp} = 0$, and we
let $\bn\cdot q = \bn\cdot q_1 - \bn\cdot q_2$ and $ q^\mu_{\perp}=
q^\mu_{1\perp} - q^\mu_{2\perp}$. In addition $K$ is the sum of
residual momenta and $k$ is the difference of residual momenta.  In
SCET there are always implicit sums over label momenta that come
with collinear fields. This turns the integrals over residual momenta
into integrals over the full momenta.  There also is an
implicit delta function which conserves label momenta. This delta
function enforces the above momentum relations (see
Ref.~\cite{Bauer:2001ct} for details).  We get
\begin{equation}
\sum_{\bn\cdot q} \sum_{q_{\perp}} \int \frac{d^4 k}{(2 \pi)^4} \to
\frac{1}{2}  \int \frac{dk^+}{2 \pi} 
\int \frac{d\bn\cdot q}{2 \pi}  \int \frac{d^2q_\perp}{(2 \pi)^2} \,.
\end{equation}
Once we make this replacement, the loop integral over the relative
momentum in Eq.~(\ref{diagram3}) can be performed giving
\begin{eqnarray}\label{diagram42}
\langle 0 | T \, {\rm Tr}\big[ B^{(0) \alpha}_\perp 
\delta_{\omega,\bnP} B^{(0) \beta}_\perp \big] (x)
{\rm Tr}\big[ B^{(0) \alpha'}_\perp \delta_{\omega',\bnP} 
B^{(0) \beta'}_\perp \big] (0) |0 \rangle
& = &    
\frac{i}{2} (g^{\alpha \alpha'}_\perp g^{\beta \beta'}_\perp+
g^{\alpha \beta'}_\perp g^{\beta \alpha'}_\perp)\,\delta_{\omega,\omega'}
\nn  \\
& & \hspace{-50ex}
\times \int\frac{d^4K}{(2\pi)^4}e^{-iK\cdot x}\,
\frac{\Gamma(\epsilon) }{8 \pi^2} 
\left(4 \pi \frac{\mu^2}{-M^2-i\delta}\right)^\epsilon 
\int^1_{-1} d \xi \, \frac{1}{[(K^+/M)(1-\xi^2)]^\epsilon} \,  
\delta_{\omega,M \xi} \,. 
\end{eqnarray}
By comparing the above result to Eq.~(\ref{cscoll1}) it is
straightforward to determine the jet function,
\begin{equation}
J_\omega(k^+,\mu) = \frac{\Gamma(\epsilon)}{8 \pi^2} 
\left(4 \pi \frac{\mu^2}{-M^2-i\delta}\right)^\epsilon 
\int^1_{-1} d \xi \, \frac{1}{[(k^+/M)(1-\xi^2)]^\epsilon} \,  
\delta_{\omega,M \xi} \,. 
\end{equation}
The decay rate is given by the imaginary part of the forward
scattering amplitude.  Taking the imaginary part of the jet function
we obtain
\begin{equation}
{\rm Im } J_\omega(k^+,\mu) =  \frac{1}{8 \pi} \Theta(k^+)
 \int^1_{-1} d \xi  \, \delta_{\omega,M \xi} \,.
\end{equation}
Substituting the above equation and the hard coefficient 
$H(\omega,\mu)$ into the imaginary part of the 
forward scattering amplitude in Eq.~(\ref{morebots}), and
summing over $\omega$ gives
\begin{eqnarray}\label{noouch}
{\rm Im} T(z) &=&
\frac{2 M}{M^2} 
\int d \ell^+ \, S(\ell^+,\mu) \, \Theta[\ell^++M(1-z)]
\, \frac{1}{8 \pi}  \int_{-1}^1 d\xi  \, H(M \xi, \mu)
\nn \\
&=&
\frac{2 M}{M^2} 
\int d \ell^+ \, S(\ell^+,\mu) \, \Theta[\ell^++M(1-z)]
\, \frac{16 \pi}{3} \bigg( \frac{4 \alpha_s(M) e e_b}{3 M} \bigg)^2 \,,
\end{eqnarray}
where the $2M/M^2$ accounts for the non-relativistic normalization of
the $\Upsilon$ state in the usoft function.  The second line is
obtained by using the tree level hard coefficient
Eq.~(\ref{lohardfun}). The integral over $\xi$ then gives only a
numerical factor of two above.  However, we will show in the next
section that once logarithms are summed the hard coefficient depends
on $\xi$ and the integral is no longer trivial. Eq.~(\ref{noouch}) is
precisely in the form given in Eq.~(\ref{nonotmore}), and it is
straightforward to read off the ${\cal H}_J$.

For the final step we first note that the usoft function may formally
be written as
\begin{equation}\label{soft}
S(\ell^+,\mu) =  \langle \Upsilon | \psi^\dagger_{\bf p} \bsigma_i \chi_{-{\bf p}}
\delta(i n\cdot \partial - \ell^+) 
\chi^\dagger_{-{\bf p}'} \bsigma_i \psi_{{\bf p}'} | \Upsilon \rangle \,.
\end{equation}
It is then possible to do the integral over $\ell^+$ in
Eq.~(\ref{noouch}), giving
\begin{equation}\label{nopenouch}
{\rm Im} T(z) =   
\langle \Upsilon | \psi^\dagger_{\bf p} \bsigma_i \chi_{-{\bf p}}
\Theta[i n\cdot \partial+M(1-z)] 
\chi^\dagger_{-{\bf p}'} \bsigma_i \psi_{{\bf p}'} | \Upsilon \rangle
\, \frac{16 \pi^2}{M} \bigg( 
\frac{32 \alpha \alpha^2_s(M) e_b^2}{27 m_b^2} \bigg).
\end{equation}
In Ref.~\cite{Rothstein:1997ac} it was shown that 
\begin{eqnarray}
\langle \Upsilon | \psi^\dagger_{\bf p} \bsigma_i \chi_{-{\bf p}}
\Theta(i n\cdot \partial+M- 2 E_\gamma) 
\chi^\dagger_{-{\bf p}'} \bsigma_i \psi_{{\bf p}'} | \Upsilon \rangle
& = & 
\nn \\
& & \hspace{-30ex} 
\Theta(\mups - 2 E_\gamma) 
\langle \Upsilon | \psi^\dagger_{\bf p} \bsigma_i \chi_{-{\bf p}}
\chi_{-{\bf p}'} \bsigma_i \psi_{{\bf p}'} | \Upsilon \rangle \,,
\end{eqnarray}
where the matrix element on the right hand side is a local operator, 
which can be determined from the $\Upsilon \to \ell^+ \ell^-$ decay rate.
Remembering that $z = 2 E_\gamma / M$ we can substitute the above
equation into Eq.~(\ref{nopenouch}) to obtain the final expression for
the imaginary part of the forward scattering amplitude
\begin{equation}\label{nofinalfsa}
{\rm Im} T(z) = \Theta(\mups - M z) 
\, \frac{16 \pi^2}{M }\Gamma_0,
\end{equation}
where $\Gamma_0$ is given in Eq.~(\ref{gamma0}).  Plugging the above
equation into Eq.~(\ref{optcthm}), gives the $z\to 1$ limit of the
tree level rate, Eq.~(\ref{LOrate}).

\section{Summing Sudakov Logarithms}

One of the powers of an effective field theory approach is the ability
to sum logarithms using the renormalization group equations. If there
exists a hierarchy of well separated scales in a process, then
logarithms of ratios of these scales arise naturally in perturbation
theory. If these terms are large enough so their product with the
perturbative expansion parameter is order one, then the original
expansion is no longer valid. By matching onto an effective theory, as
we have done here, the large scale is removed, and replaced with a
running scale $\mu$.  While the matching, which determines the Wilson
coefficient, is carried out at the high scale, operators are run from
the matching scale to the low scale using the RGEs. This sums all
large logarithms into an overall factor, and any logarithms that arise
in a perturbative expansion of the effective theory are order one.

As discussed previously, near the endpoint of
$\Upsilon \to X  \gamma$ there arises a hierarchy of scales. 
As an explicit example consider the color-octet
${}^1S_0$ processes. We studied the endpoint behavior of this
contribution in a previous work~\cite{Bauer:2001rh}. At
next-to-leading order in $\alpha_s$ the Wilson coefficient in the $z
\to 1$ limit is
\begin{eqnarray}
\label{WCoct1}
C_{\bf 8}^{(1)}(^1S_0)(z) &=& 
  \frac{\alpha_s}{2\pi}\tilde{C}_{\bf 8}^{(0)}(^1S_0)
   \left[-2C_A \left(\frac{\log(1-z)}{1-z}\right)_+ - 
   \left(\frac{23}{6} C_A - \frac{n_f}3\right)\left(\frac1{1-z}\right)_+
   \right] ,
\end{eqnarray}
where $\tilde{C}_{\bf 8}^{(0)}(^1S_0) = 16\alpha_s\alpha
e_b^2\pi/M^2$.  If this coefficient is integrated over the endpoint
region ($1-v^2 < z < 1$), the first plus distribution on the
right-hand side gives rise to a double logarithm, $\log^2 v^2$, and
the second plus distribution gives a single logarithm, $\log v^2$.
Both of these are numerically of order $1/\alpha_s$. This clearly
ruins the perturbative expansion. We showed in
Ref.~~\cite{Bauer:2001rh} that these large logarithms are summed by
matching onto SCET at the scale $M$ and running to the usoft
scale. The is also true of the color-octet ${}^3P_J$ contribution.

For the color-singlet ${}^3S_1$ contribution we can not 
explicitly give the NLO contribution in the endpoint region
since this has only been calculated numerically~\cite{Kramer:1999bf}.
However, while the NLO correction to the color-singlet 
contribution does not change the LO prediction very much for 
low and intermediate values of $z$, it does change
the endpoint prediction by an amount of order one. This, as we 
will show, is due to the
presence of endpoint logarithms. However,  unlike 
the color-octet case these endpoint logarithms are
single, {\it not} double logarithms. 

\begin{figure}[t]
\centerline{\includegraphics[width=2.5in]{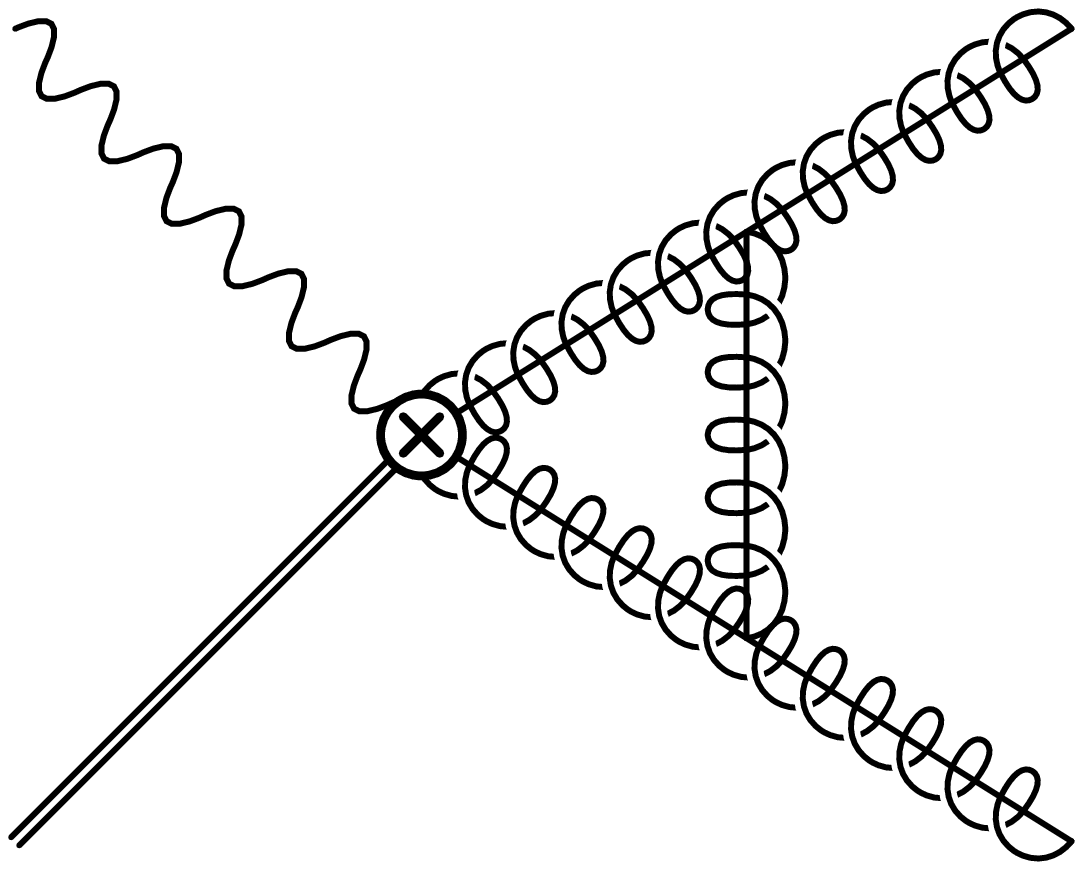}\qquad
\includegraphics[width=2.5in]{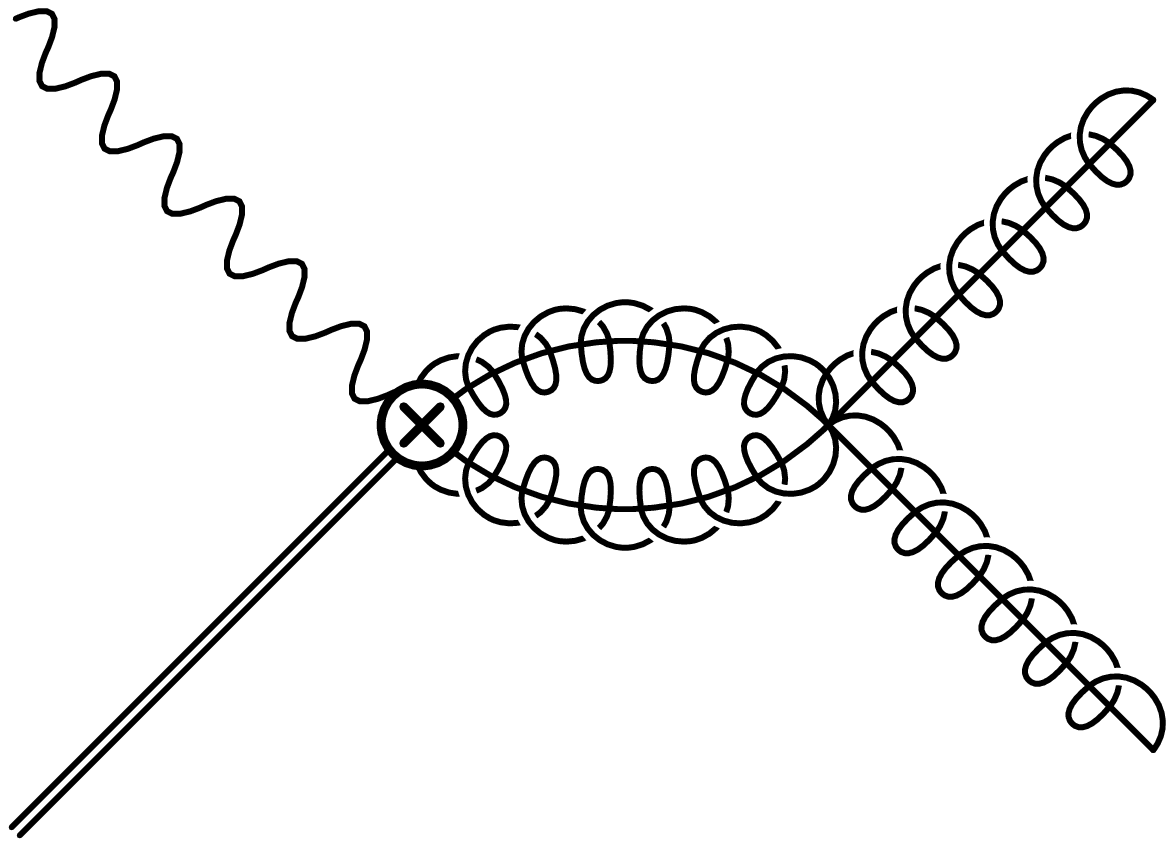}}
\centerline{\includegraphics[width=2.5in]{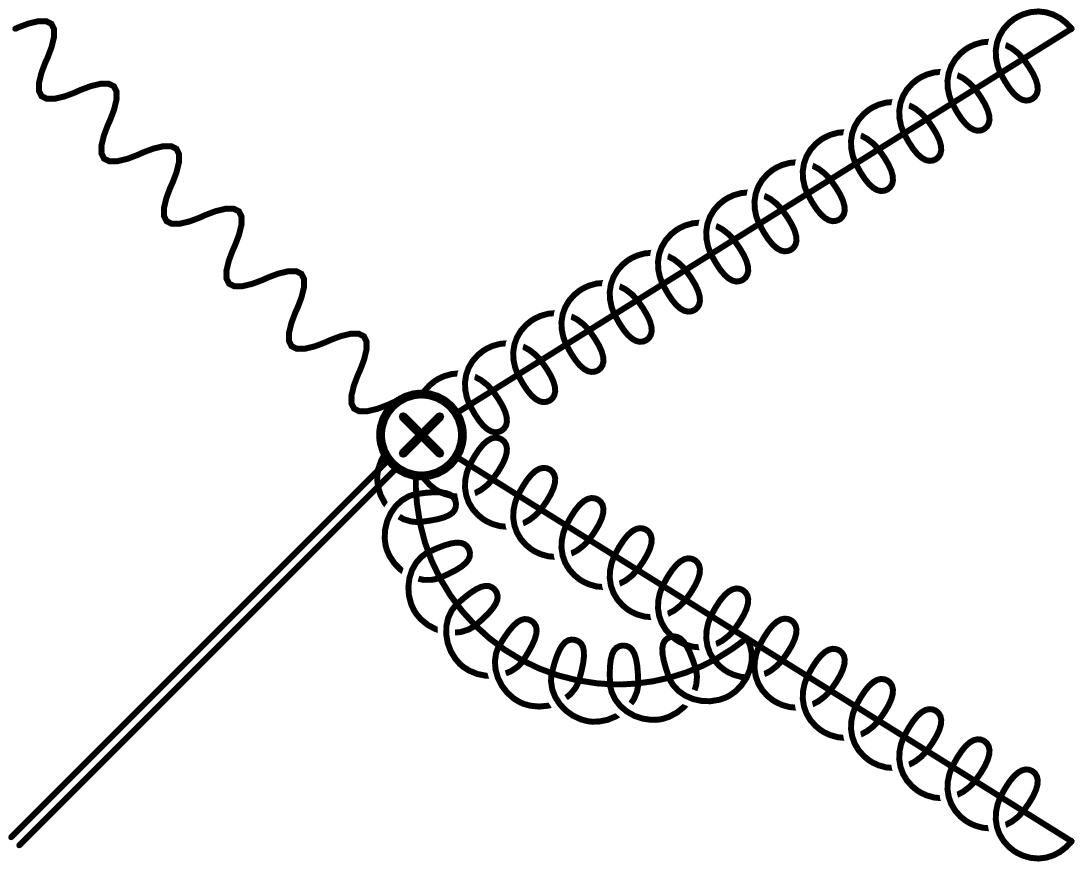}\qquad
\includegraphics[width=2.5in]{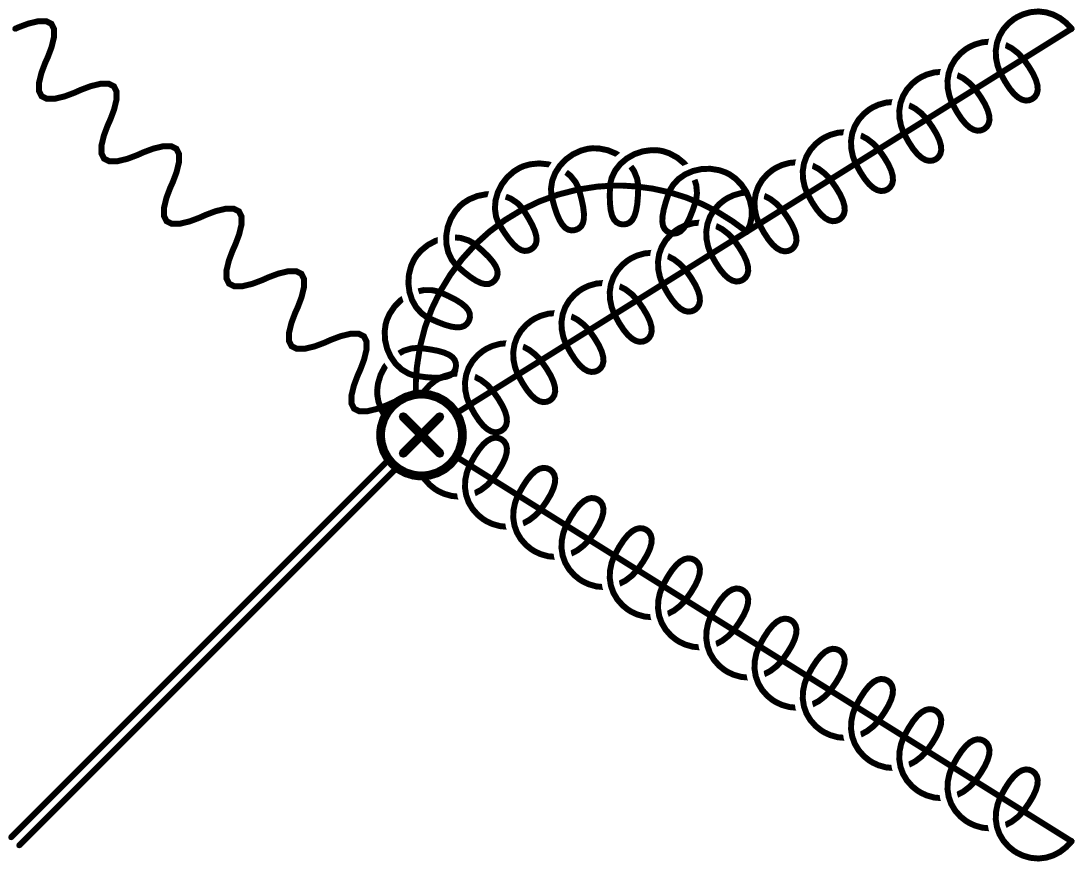}}
\caption{\it Diagrams needed to calculate the counterterm to 
the color-singlet operator.
\label{oneloop}}
\end{figure}

In Section~\ref{matchsection} we matched onto the SCET color-singlet
operator. This integrates out the scale $M$. We now run the
color-singlet operator from the hard scale to the collinear scale,
which sums all logarithms of $1-z$.  Unlike the color-octet
contribution the color-singlet operator does not run below the
collinear scale. To run the color-singlet operator given in
Eq.~(\ref{3s1op2}), we calculate the counterterm for the operator,
then determine the anomalous dimension, and finally use this in the
RGEs. The four graphs that need to be evaluated to obtain the
counterterm are shown in Fig.~\ref{oneloop}.
Note that the diagrams only involve collinear gluons. All of the
diagrams involving usoft gluons vanish because
the color-singlet process is transparent to usoft gluons. This
is clear from  the previous section where we showed that
usoft Wilson lines completely cancel in the color-singlet 
contribution. This is very different than the color-octet
contribution where usoft gluons play a crucial role. The Feynman rules
for the vertex operators are given in Appendix~\ref{appFR}.  The
result of each of the loop integrals is complicated so we do not give
each individual expression. However, when all the terms are added we
obtain a relatively simple result for the one-loop UV-divergent term
\begin{eqnarray}\label{UVterms}
{\cal A} &=& 
\frac{1}{\epsilon} \, 
\chi^\dagger_{-{\bf p}} \Lambda\cdot\bsigma^\delta \psi_{\bf p}
\sum_{\omega }  
  \Gamma^{(1,{}^3S_1)}_{\alpha \beta \delta \mu} ( M, \omega)
{\rm Tr} \big\{ B^\alpha_\perp \, \delta_{ \omega,\cP_{-}} 
  \, B^\beta_\perp \big\} 
\\
&& \qquad\qquad \times \frac{\alpha_s(\mu) C_A}{2 \pi} \bigg[ 1+
\frac{M^2+ \omega^2 }{M^2}
 \bigg(\frac{M}{M+\omega} \ln\frac{M-\omega} {2M}
 + \frac{M}{M-\omega} \ln\frac{M+\omega}{2M}\bigg) \bigg] ,\nn
\end{eqnarray}
where we have introduced a sum over the index $\omega$.
Note that the divergent contribution depends on the large momentum
component of the collinear gluons. This means that the counterterm
will depend on the large momentum components, as will the anomalous
dimension.  This divergent piece must be canceled by 
$Z_3 / Z_{\cal O}-1$, where $Z_{\cal O}$ is the counterterm for the
color-singlet vertex in SCET, and $Z_3$ is the gluon wave function
counterterm
\begin{equation}
Z_3 = 1 + \frac{\alpha_s}{4 \pi} \frac{1}{\epsilon} 
\left( C_A \frac{5}{3} -n_f \frac{2}{3} \right) \,.
\end{equation}
This leads to
\begin{equation}
Z_{\cal O}-1 = 
\frac{1}{\epsilon} \, \frac{\alpha_s}{2 \pi} \bigg\{ C_A \bigg[ \frac{11}{6} +
\frac{M^2+\omega^2 }{M^2}
 \bigg(\frac{M}{M+\omega} \ln\frac{M-\omega}{2M} 
  + \frac{M}{M-\omega} \ln\frac{M+\omega}{2M}\bigg) \bigg] 
  - \frac{n_f}{3}\bigg\} \,.
\end{equation}
The anomalous dimension is obtained through the standard method, and
the RGE for the color-singlet Wilson coefficient is
\begin{equation}
\mu \frac{d}{d \mu} \Gamma^{(1,{}^3S_1)}(\mu, \omega) = 
\gamma(\mu, \omega) \Gamma^{(1,{}^3S_1)}(\mu, \omega) \,.
\end{equation}
We have made the dependence on the momentum label $\omega$ 
explicit to emphasize that the anomalous dimension depends
on these labels. Solving the RGE gives
\begin{eqnarray}\label{resumcoeff}
\ln \! \bigg( \frac{\Gamma^{(1,{}^3S_1)}(\mu, \omega)}
    {\Gamma^{(1,{}^3S_1)}(M, \omega)} \bigg) &=& 
 \\
& & \hspace{-20ex}
\frac{2}{\beta_0} \bigg\{ C_A 
\bigg[ \frac{11}{6} + \frac{M^2+ \omega^2 }{M^2}
\bigg(\frac{M}{M+\omega} \ln\frac{M-\omega}{2M} 
  + \frac{M}{M-\omega} \ln\frac{M+\omega}{2M}\bigg) \bigg] 
 - \frac{n_f}{3}\bigg\}  \ln\!\bigg( \frac{\alpha_s(\mu)}{\alpha_s(M)}\bigg),
\nn
\end{eqnarray}
where $\Gamma^{(1,{}^3S_1)}(M, \omega)$ is given in
Eq.~(\ref{matching1}).  Logarithms of the form $\ln(\mu/M)$ have been
summed into $\Gamma^{(1,{}^3S_1)}(\mu, \omega)$, and any logarithms in
the operator are of the form $\ln(\mu_c/\mu)$, where $\mu_c \approx
M\sqrt{1-z}$ is the collinear scale.  If we take $\mu\sim\mu_c$
logarithms in the operator will be small, and all large logarithms of
the ratio $\mu_c/M$ will sit in the Wilson coefficient.

We can now obtain the resummed rate, by substituting the Wilson
coefficient at the collinear scale $\mu = M\sqrt{1-z}$, 
Eq.~(\ref{resumcoeff}), into Eq.~(\ref{noouch}), giving 
\begin{eqnarray}\label{ouch}
{\rm Im} T(z) &=&  2 M \int d \ell^+ \, S(\ell^+,\mu) \, \Theta[\ell^++M(1-z)]
\, \frac{16 \pi}{3} \bigg( \frac{4 \alpha_s(M) e e_b}{3 M^2} \bigg)^2 
\\
& & \times 
\int_0^1 d \eta \bigg[ \frac{\alpha_s(M\sqrt{1-z})}{\alpha_s(M)} 
\bigg]^{2\gamma(\eta)},
\nn
\end{eqnarray}
where 
\begin{equation}\label{anom}
\gamma(\eta) \equiv \frac{2}{\beta_0} \bigg\{ C_A \bigg[ \frac{11}{6}
+\big(\eta^2 + (1-\eta)^2 \big) \bigg( \frac{1}{1-\eta} \ln \eta
+ \frac{1}{\eta}\ln (1-\eta) \bigg) \bigg] -\frac{n_f}{3} \bigg\}.
\end{equation}
We have substituted $\xi = 2\eta - 1$.  As in the
previous section, using the formal result for the usoft function,
Eq.~(\ref{soft}), and the results of Ref.~\cite{Rothstein:1997ac} we
obtain the final expression for the imaginary part of the forward 
scattering amplitude including resummation
\begin{equation}\label{finalfsa}
{\rm Im} T(z) = \Theta(\mups - M z) 
\, \frac{16 \pi^2}{M }\Gamma_0 
\int_0^1 d \eta \bigg[ \frac{\alpha_s(M\sqrt{1-z})}{\alpha_s(M)} 
\bigg]^{ 2\gamma(\eta) } \,,
\end{equation}
where $\Gamma_0$ is given in Eq.~(\ref{gamma0}). The resummed 
color-singlet contribution to the decay rate in the endpoint region is
obtained by inserting the above equation into Eq.~(\ref{optcthm})
\begin{equation}\label{resrate}
\frac1{\Gamma_0}\frac{d\Gamma_{\rm resum}}{dz} =
\Theta(\mups - M z) \, z 
\int_0^1 d \eta \bigg[ \frac{\alpha_s(M\sqrt{1-z})}{\alpha_s(M)}
\bigg]^{ 2\gamma(\eta) } .
\end{equation}
Note we can expand Eq.~(\ref{resrate}) in powers of $\alpha_s(M)$
and obtain an analytic expression for the NLO logarithmic
contribution
\begin{equation}\label{expresrate}
\frac{1}{\Gamma_0} \frac{d \Gamma}{d z} = 
z \bigg\{ 1 + \frac{ \alpha_s(M)}{6 \pi} \big[ C_A (2 \pi^2 - 17)+2 n_f 
\big] \ln(1-z) \bigg\} \,.
\end{equation}
As $z$ approaches one the ${\cal O}(\alpha_s)$ term becomes negative
and of order one, precisely the behavior observed in
Ref.~\cite{Kramer:1999bf} for the NLO decay rate at the endpoint.

\section{Phenomenology and Discussion} 

In this section we combine the different contributions to obtain  a 
prediction for the photon spectrum in $\Upsilon\to X\gamma$ decay.
We will marry our expression for the color-singlet spectrum in the 
endpoint with the leading order result by interpolating between the 
two
\begin{equation}\label{fulleq}
\frac{1}{\Gamma_0} \frac{d \Gamma_{\rm int}}{d z}= 
\bigg( \frac{1}{\Gamma_0} \frac{d \Gamma_{\rm LO}^{\rm dir}}{d z} - z \bigg)
+ \frac{1}{\Gamma_0} \frac{d \Gamma_{\rm resum}}{d z} \,.
\end{equation}
The term
in brackets vanishes as $z \to 1$, leaving only the
resummed contribution in that region. Away from the endpoint the
resummed contribution combines with the $-z$ to give higher order in
$\alpha_s(M)$ corrections.  This clear from Eq.~(\ref{expresrate}).

Before we proceed we need the NRQCD matrix
elements.  The color-singlet matrix element can be related to the
wavefunction at the origin, Eq.~(\ref{singletWF}), which can be
calculated in potential models, on the lattice, or extracted from the
$\Upsilon$ leptonic decay rate,
\begin{equation}
\Gamma(\Upsilon \to e^+ e^-) =
\frac{2\pi\alpha^2 e_b^2}{3}
\left[ 1 - \frac83 \frac{\alpha_s(m_b)}{\pi} 
  - \frac13 \frac{M_{\Upsilon(nS)} - 2 m_b}{2 m_b} \right]^2
\frac{\langle\Upsilon\vert{\cal O}_1(^3S_1)\vert\Upsilon\rangle}{m_b^2}.
\label{MEextraction}
\end{equation}
The relativistic correction was first expressed in terms of
$M_\Upsilon - 2 m_b$ by Gremm and Kapustin \cite{Gremm:1997dq}.
The matrix element is sensitive to the value of the $b$ quark mass.
We will use $m_b = 4.8$ GeV, and $\alpha_s(m_b) = 0.21$, giving  
$\langle\Upsilon\vert{\cal O}_1(^3S_1)\vert\Upsilon\rangle = 3.40
{\rm\ GeV}^3$.

The color-octet matrix elements are more difficult to determine.  They
can, in principle, be extracted from data or calculated on the lattice
(e.g., see Ref.~\cite{Bodwin:2001mk}).  NRQCD predicts that the
color-octet matrix elements scale as $v^4$ compared to the
color-singlet matrix element.  In Ref.~\cite{Petrelli:1998ge} it was
argued that a factor of $1/2 N_c$ should be included in a naive
estimate of the color-octet matrix elements.  For now, we leave these
parameters free.

For the color-octet $^1S_0$ and $^3P_0$ rates, we need a structure
 function \cite{Rothstein:1997ac,Wolf:2001pm}.  We will use the simple model
introduced in Ref.~\cite{Bauer:2001rh}
\begin{eqnarray}
f(k^+) &=& N \left( 1-\frac{k^+ + \Lambda_1}{\bar \Lambda} \right)^{a} \,
         e^{(1+a)(k^+ + \Lambda_1)/\bar \Lambda} \;,
\label{shape1}
\end{eqnarray}
where $N$ is chosen so that the integral of the structure function is
normalized to one. In principle the structure function can be
different for the different color-octet states. But since we are
ignorant of the non-perturbative structure function, we will naively
use the same model for both the ${}^1S_0$ and ${}^3P_0$
configurations.  For quarkonium the first moment of the structure function
with respect to $k^+$ is
\begin{eqnarray}
\Lambda_1 =
\frac{ \langle\Upsilon|\sum_{{\bf p}, {\bf p'}} [\psi^\dagger_{\bf p'}
T^a \Gamma_i \chi_{-{\bf p'}}]  i D^+
[\chi^\dagger_{-{\bf p}} T^a \Gamma_i \psi_{\bf p}]|\Upsilon\rangle }
{\langle\Upsilon| \sum_{{\bf p}, {\bf p'}} [\psi^\dagger_{\bf p'}
T^a \Gamma_i \chi_{-{\bf p'}}][ \chi^\dagger_{-{\bf p}} T^a
\Gamma_i \psi_{\bf p}]|\Upsilon\rangle }\,.
\end{eqnarray}
The integration limits for $k^+$ are from $-M$
to $M_\Upsilon-M$, and both $\bar \Lambda$ and $\Lambda_1$ are
non-perturbative parameters related through $\bar{\Lambda}=M_\Upsilon
-M -\Lambda_1$. We use the following numbers in our plots: $a=1$,
$\bar \Lambda = 480\, {\rm MeV}$, and $\Lambda_1 = -620\, {\rm MeV}$.

\begin{figure}[t]
\centerline{\includegraphics[width=5in]{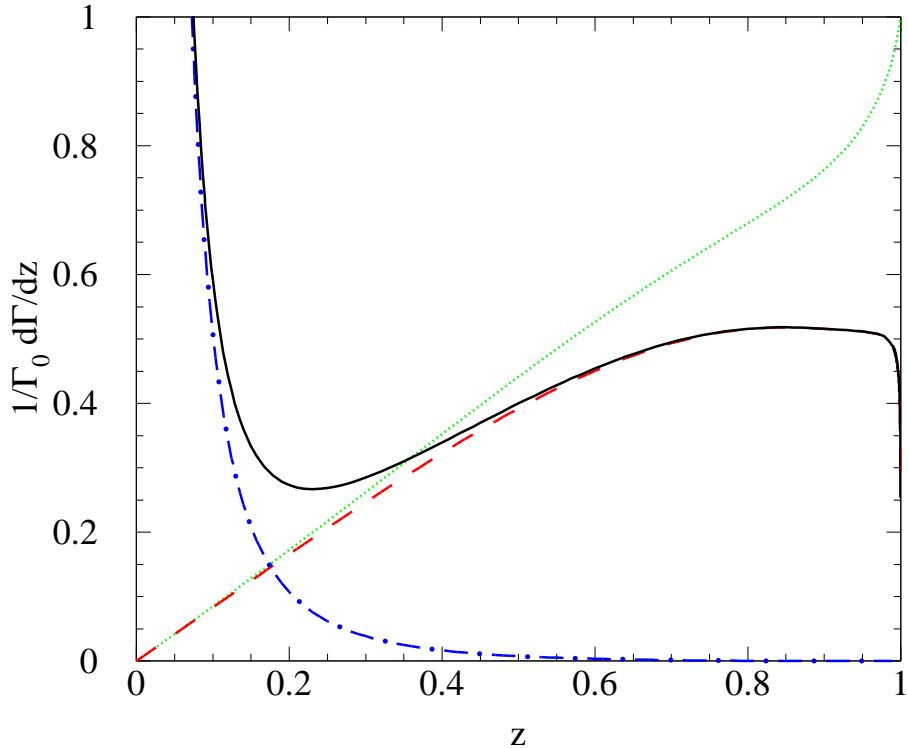}}
\caption{\it The color-singlet rate.  The dotted curve is the
tree-level direct rate.  The dashed curve is the interpolated resummed direct
rate.  The dot-dashed curve is the fragmentation contribution, and the
solid curve is the sum of the resummed rate and the fragmentation rates.}
\label{CSfig}
\end{figure}

In Fig.~\ref{CSfig} we show the color-singlet rate.  
Since each contribution is proportional to the same matrix element, we
can normalize the rate to $\Gamma_0$ in Eq.~(\ref{gamma0}).  The
dotted curve is the tree-level direct rate, Eq.~(\ref{LOrate}), 
the dashed is the interpolated resummed result, Eq.~(\ref{fulleq}), 
the dot-dashed is the fragmentation contribution,
Eq.~(\ref{CSfrag}), and the solid is the sum of the interpolated 
resummed and fragmentation contributions.  As can be seen, the 
resummed rate rate turns over and decreases near the endpoint.  
At some point the resummed rate has a problem with the Landau pole, 
which can be handled in a manner described in Ref.~\cite{Leibovich:2001ra}; 
however the Landau pole is not reached until $z\sim0.999$, at 
which point the curve is in the resonance region and our results are 
no longer valid.

\begin{figure}[t]
\centerline{\includegraphics[width=5in]{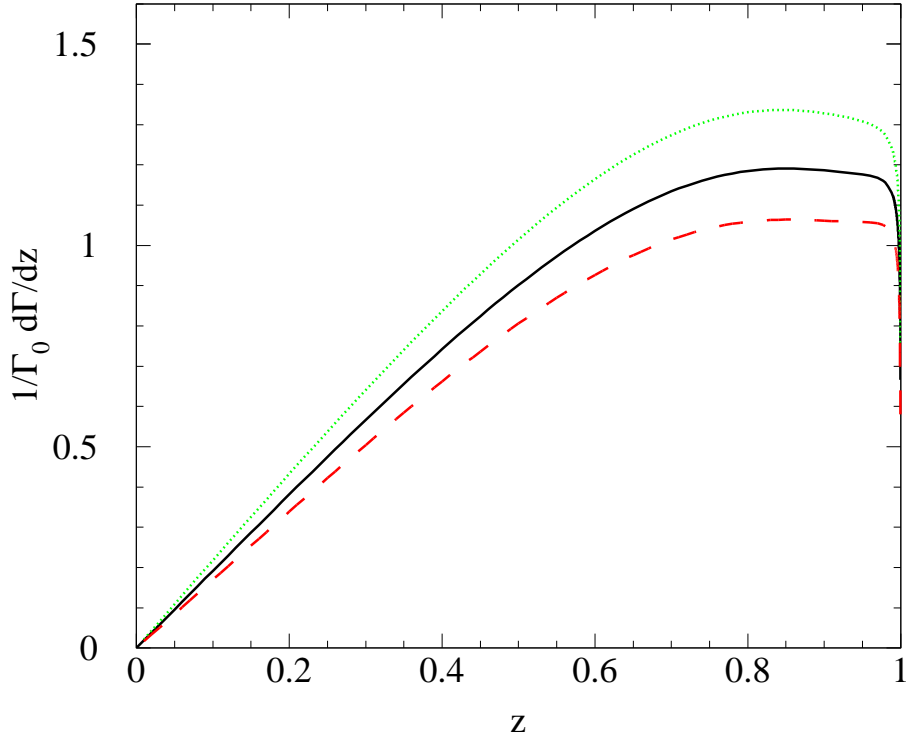}}
\caption{\it The resummed color-singlet rate for different choices of
collinear scale $\mu_c$.  The solid curve uses $\mu_c = M\sqrt{1-z}$,
the dotted $\mu_c=M\sqrt{2(1-z)}$, and the dashed $\mu_c = M\sqrt{(1-z)/2}$.}
\label{CSscale}
\end{figure}

The resummed result depends on the collinear scale.  In
Fig.~\ref{CSfig} the collinear scale was set to $\mu_c = M\sqrt{1-z}$.  In
Fig.~\ref{CSscale} the solid curve again uses this choice of $\mu_c$.
The dotted is made with the collinear scale set to $\mu_c =
M\sqrt{2(1-z)}$, and the dashed has $\mu_c = M\sqrt{(1-z)/2}$.  The
choice of scale is a higher order effect, which can only be determined
by calculating higher order corrections to our results.  The scale
variation changes the rate by about 10\%.  This can be considered part
of the theoretical uncertainty.

\begin{figure}[t]
\centerline{\includegraphics[width=5in]{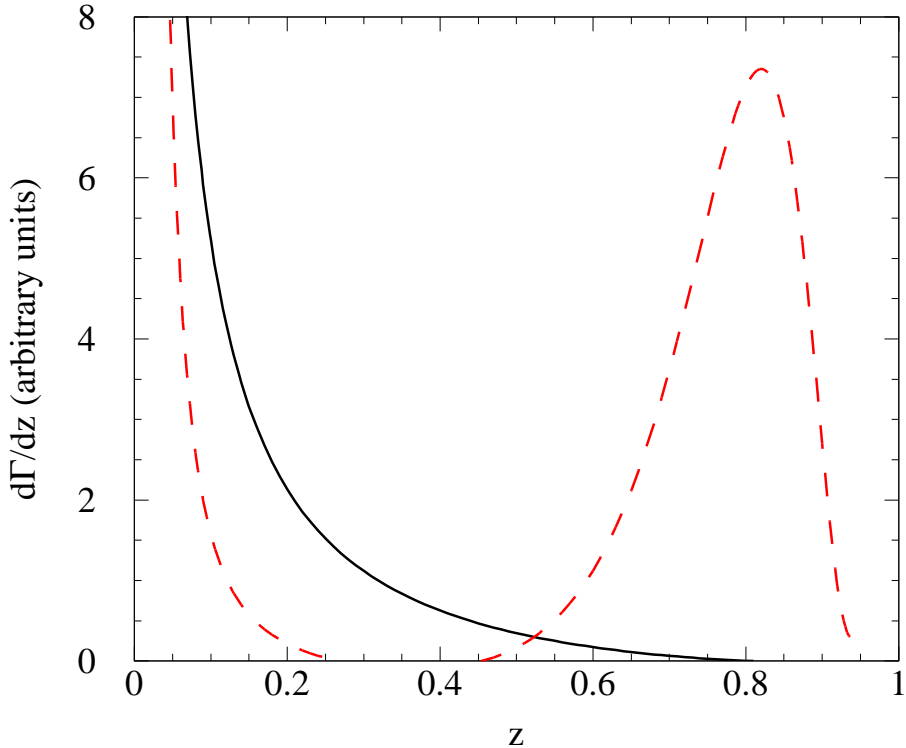}}
\caption{\it The color-octet rate with arbitrary choice of
NRQCD matrix elements.  The solid curve is the octet $^3S_1$
fragmentation contribution, while the dashed curve is the octet
$^1S_0$ and $^3P_0$ resummed rate convoluted with the shape function
and the octet $^1S_0$ and $^3P_0$ fragmentation rate.}
\label{COfig}
\end{figure}

We also need to include the color-octet contributions to the decay
rate, which requires knowledge of the the color-octet matrix elements.
Since these are unknown at present, we first show in Fig.~\ref{COfig}
the color-octet contributions with arbitrary choices of 
$\langle\Upsilon\vert{\cal O}_8(^3S_1)\vert\Upsilon\rangle$ and 
\begin{equation}
\langle\Upsilon\vert{\cal O}_8(^1S_0)\vert\Upsilon\rangle +
7\frac{\langle\Upsilon\vert{\cal O}_8(^3P_0)\vert\Upsilon\rangle}{m_b^2},
\label{octcomb}
\end{equation}
the linear combination which occurs in both the fragmentation and the
resummed rates.  The solid curve is the octet $^3S_1$ fragmentation
contribution, and the dashed curve is the octet $^1S_0$ and $^3P_0$
resummed rate convoluted with the shape function added to 
the octet $^1S_0$ and $^3P_0$ fragmentation rate.
The $^3S_1$ octet fragmentation rate has support at larger values of
$z$ than the other fragmentation contributions at this order.
However, higher order perturbative corrections to the other octet
fragmentation contributions do increase the support at higher $z$
\cite{Maltoni:1999nh}.  The resummed octet rate produces a large peak
near $z\sim0.8$ which does not appear to be present in the data.  We
will therefore have to choose the combination in Eq.~(\ref{octcomb})
to be small enough to agree with  data. 

The CLEO collaboration has measured the number of photons in inclusive
$\Upsilon(1S)$ radiative decays in Ref.~\cite{Nemati:1996xy}.  The
data presented does not remove the efficiency or energy resolution and is
the number of photons in the barrel region of the detector, $|\cos\theta|<0.7$.
Therefore, in order to compare our theoretical prediction to the data,
we first integrate over the barrel region and then convolute with the
efficiency that was modeled in the CLEO paper.  We did not do a
bin-to-bin smearing of our prediction.  Since we
do not know the size of the color-octet matrix elements, we will set
the $^1S_0$ and $^3P_0$ matrix elements to zero, and the $^3S_1$
matrix element to
\begin{equation}
\langle\Upsilon\vert{\cal O}_8(^3S_1)\vert\Upsilon\rangle =
v^4 \langle\Upsilon\vert{\cal O}_1(^3S_1)\vert\Upsilon\rangle,
\end{equation}
where we set $v^2 = 0.08$.
\begin{figure}[t]
\centerline{\includegraphics[width=5in]{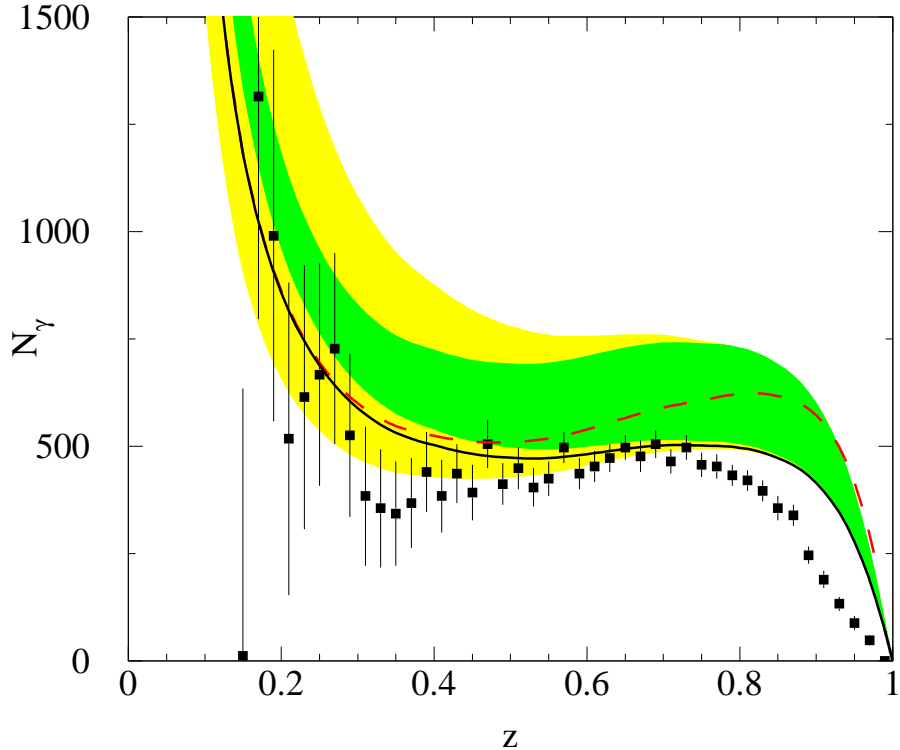}}
\caption{\it The inclusive radiative $\Upsilon$ photon spectrum,
compared with data from CLEO \cite{Nemati:1996xy}.  The error bars on
the data are statistical only.  The theoretical prediction, shown by
the solid curve, includes the color-singlet interpolated resummed and
fragmentation contribution and the color-octet $^3S_1$
fragmentation contribution, with
$\alpha_s$ set to the value extracted in \cite{Nemati:1996xy}. The
other color-octet matrix elements are set to zero.  The dashed line is
the direct tree-level and fragmentation result, with the same choice
of $\alpha_s$.  The theoretical prediction using the PDG value for
$\alpha_s$, shown by the shaded bands, includes the
color-singlet interpolated resummed plus fragmentation contribution 
and the color-octet $^3S_1$ fragmentation contribution.
The inner band is obtained by varying
$m_b$, $\alpha_s$, and the collinear scale, 
while the outer band includes variation of the
fragmentation function, as described in the paper.}
\label{comparedata}
\end{figure}

In Fig.~\ref{comparedata} we compare our prediction to the data.  The
error bars on the data are statistical only.  The dashed line is the
direct tree-level and fragmentation result,  and the solid curve is the sum of 
the interpolated resummed result and the fragmentation result.
For these two curves we used the value of $\alpha_s$ extracted
by CLEO from these data, $\alpha_s(M_\Upsilon) = 0.163$, which corresponds to
$\alpha_s(M_Z) = 0.110$ \cite{Nemati:1996xy}.  
While not a perfect fit, the shape of the
resummed result is closer to the data than the tree-level curve.  We
also show in this plot the interpolated 
resummed and fragmentation result, using
the PDG value of $\alpha_s(M_Z)$, including theoretical uncertainties,
denoted by the shaded region.  To obtain the darker band, we first
varied the choice of $m_b$ between $4.7 {\rm\ GeV} < m_b < 4.9 {\rm\
GeV}$ and the value of $\alpha_s$ within the errors given in the PDG,
$\alpha_s(M_Z) = 0.1172\pm0.002$ \cite{Hagiwara:pw}.  Varying $m_b$
and $\alpha_s$ modifies the extraction of the color-singlet matrix
element in Eq.~(\ref{MEextraction}) from $3.31 {\rm\ GeV}^3$ to $3.56
{\rm\ GeV}^3$.  We also varied the collinear scale, $\mu_c$ as we did
in Fig.~\ref{CSscale} from $M\sqrt{(1-z)/2} < \mu_c < M\sqrt{2(1-z)}$.
Finally, the lighter band also includes the variation, within the
errors, of the parameters for the quark to photon fragmentation
function, Eq.~(\ref{quarkfrag}), that was extracted by ALEPH
\cite{Buskulic:1995au}.  The low $z$ prediction is dominated by the
quark to photon fragmentation coming from the color-octet $^3S_1$
channel.  We did not assign any error to the color-octet $^3S_1$
matrix element.  Since it is unknown, there is a very large
uncertainty in the lower part of the prediction that we decided not to
show. We also do not include in our error estimate contributions 
from higher order SCET operators, which are generically suppressed 
by a least a power of $\lambda/M \sim \sqrt{1-z}$.
Note that the color-octet ${}^1S_0$ and ${}^3P_0$ contribution
increases the theoretical prediction at the upper endpoint, as shown in
Fig.~\ref{COfig}. It is thus clear the data favors a very small value for
the linear combination Eq.~(\ref{octcomb}). This is why we set these
matrix elements to zero in our analysis.  If we used small negative values
for these matrix elements we would fit the shape in the endpoint
region a bit better. 

As mentioned earlier, the NLO calculation of the photon spectrum was
previously calculated numerically in Ref.~\cite{Kramer:1999bf}.  The
$\alpha_s$ correction is small for low and intermediate values of
$z$, but becomes larger above $z\approx 0.8$ until it is of order one 
above $z \approx 0.9$. Much of the difference between the LO and NLO 
piece in  the endpoint is due to a  $\alpha_s\log(1-z)$  term. However,
some of the difference is due to the non-logarithmic contributions. Here
we resummed logarithms of the form $\alpha_s\log(1-z)$ so our
result includes the logarithmic piece of the NLO contribution. However, 
we do not incorporate any of the non-logarithmic NLO corrections.
This would be done by matching onto SCET at NLO. Since some of the
difference between the shape of the LO and NLO contribution at the endpoint
is due to these terms, a NLO matching may give a better fit to data in 
the endpoint region.

The logarithms in the color-singlet rate have previously been studied
in Refs.~\cite{Photiadis:1985hn,Hautmann:2001yz}.  In 
Ref.~\cite{Photiadis:1985hn} collinear logarithms were summed 
including mixing with quarks. Since we neglected mixing with 
quark operators (see footnote~\ref{noquarkfoot}) we compare to
the expression obtained in Ref.~\cite{Photiadis:1985hn}  by setting 
the quark contribution zero. When we do that we find that the 
results are in agreement. In Ref.~\cite{Hautmann:2001yz} it was 
argued that all logarithms cancel in the color-singlet rate, which is
not what we found. However, the analysis of Ref.~\cite{Hautmann:2001yz} 
was based on the eikonal approximation, which is equivalent to including
only usoft gluon modes. It did not include purely collinear modes. We do 
agree with the statement in Ref.~\cite{Hautmann:2001yz} that all
logarithms due to usoft gluons cancel in the rate.  This is clear in 
Section IV where we found that at leading order in $\lambda$ the 
usoft Wilson lines cancel in the usoft function. Equivalently we saw 
this cancellation in Section VI where we found that here are no diagrams
renormalizing the color-singlet operator involving usoft gluons.
However, there are logarithms due to collinear gluons.  Note that at
higher order in $\lambda$, there would be diagrams with usoft gluons
renomalizing the operator.  This would correspond to logarithms in
the derivative of the rate, whose existence was pointed out in
Ref.~\cite{Hautmann:2001yz}.

In Ref.~\cite{Field:cy} a model is introduced based on the idea
that the outgoing gluons shower to produce a jet of particles with a
non-zero invariant mass. Specifically in this model gluons are given 
a mass of order $m_g\sim 1 {\rm\ GeV}$, which is an estimate based
on a Monte Carlo calculation of the invariant mass of the final state jet.  
It is found that this approach gives a very good fit to the 
data~\cite{Field:2001iu}.  If, as was discussed at the beginning of Section V, 
we treat the collinear scale $M\sqrt{1-z}$ as non-perturbative, we would also
be forced to introduce a model for the jet and usoft functions (since they would
be non-perturbative). This approach would then be equivalent to the approach
in Refs.~\cite{ Field:cy,Field:2001iu}, if we used as a model for the jet 
function the LO calculation carried out in Section V including a gluon mass
of order $m_g\sim 1 {\rm\ GeV}$.

\section{conclusion}

The photon energy spectrum in $\Upsilon\to X\gamma$ decay proves to be
far richer than the simple leading order prediction based upon the
color-singlet model. As was first pointed out in
Ref.~\cite{Catani:1995iz}, at low photon energies it is crucial to
include fragmentation contributions.  In addition it has long been
known that at high photon energies usoft gluon effects must be
incorporated. In Ref.~\cite{Field:cy} a model was introduced to
accomplish this, however, the result is no longer a prediction of QCD
alone. In this paper we systematically treat the endpoint region in
the soft-collinear effective theory.  Because SCET is constructed to
reproduce QCD in the endpoint region the results of our calculation
are model independent predictions of QCD.

Specifically we have determined the behavior of the color-singlet
contribution at the endpoint.  This calculation consists of matching
onto a color-singlet operator in SCET, and running the operator from
the matching scale to the collinear scale. The first step integrates
out the hard scale set by the $\Upsilon$ mass, and the second step
sums logarithms of the ratio of the hard and collinear scales. Our
result for the logarithms agrees with Ref.~\cite{Photiadis:1985hn}.
Finally we obtain the decay rate by performing an OPE and matching
onto a color-singlet shape function. As pointed out by Rothstein and
Wise~\cite{Rothstein:1997ac} this shape function can be calculated.

We combine the color-singlet contribution with the color-octet
contribution calculated in Ref.~\cite{Bauer:2001rh} to obtain a
complete theoretical prediction for the photon spectrum at the
endpoint. Furthermore we combine our result with an NRQCD calculation
that includes the fragmentation contribution at low photon energy to
obtain a theoretical prediction for the entire photon spectrum. We
compare our calculation to data taken by the CLEO
collaboration~\cite{Nemati:1996xy}. Overall there is good agreement
with the data. Moreover the results at the endpoint are now in much
better agreement with data than they were previously with the leading
order NRQCD prediction. However there is still a discrepancy between
data and theory, and in the size and shape of the spectrum.  This
discrepancy might be corrected by including contributions that are
higher order in strong coupling $\alpha_s$ and the SCET power 
counting $\lambda$.

\acknowledgments 
We would like to thank Christian Bauer, David Besson, Roy Briere, Ira
Rothstein, and Iain Stewart for helpful discussions. In addition
S.F. would like to thank the two other members of the groove lounge,
Matt Martin, and Hael Collins for useful discussion.  This work was
supported in part by the Department of Energy under grant numbers
DOE-ER-40682-143 and DE-AC02-76CH03000.

\appendix

\section{SCET operators and Feynman rules}\label{appFR}

The most general ${}^3P_J$ operator is given by
\begin{equation} \label{3pjop1}
   \chi^\dagger_{-{\bf p}} \Gamma_{\alpha \mu \sigma \delta}(\bnP) 
   B^\alpha_\perp  \Lambda \cdot \widehat{{\bf p}}^\sigma 
   \Lambda \cdot \bsigma^\delta \psi_{\bf p} \,,
\end{equation}
where  the tree level Feynman rule for this operator is
\begin{equation}\label{3pjfeynrule}
\Gamma_{\alpha \mu \sigma \delta}(\omega) 
\Big( \epsilon^\perp_\alpha- \frac{q^\perp_\alpha}{\bn\cdot q} 
      \bn\cdot\epsilon \Big)
\Lambda \cdot \widehat{{\bf p}}^\sigma \eta^\dagger_{-{\bf p}}
 \Lambda \cdot \bsigma^\delta
T^A \xi_{\bf p}\,,
\end{equation}
and $\widehat{{\bf p}} = p/M$.  To match we expand the tree level QCD
amplitude to leading order in $\lambda$:
\begin{equation}\label{1s0treeQCD2}
4 g_s e e_b  
(g^{\perp}_{\alpha \delta} g^{\perp}_{\mu \sigma} 
+ g^{\perp}_{\alpha \sigma} g^{\perp}_{\mu \delta} 
+ g^{\perp}_{\alpha \sigma} n^\delta \bn^\mu 
- g^{\perp}_{\alpha \mu} n^\delta n^\sigma)
\Big( \epsilon^\perp_\alpha- \frac{q^\perp_\alpha}{\bn\cdot q} 
      \bn\cdot\epsilon \Big)
\Lambda \cdot \widehat{{\bf p}}^\sigma \eta^\dagger_{-{\bf p}}
\Lambda \cdot \bsigma^\delta
T^A \xi_{\bf p}\,.
\end{equation}
If we take the photon to be real ({\it i.e.} to only have $\perp$
polarization) then the $\bn^\mu$ term vanishes and the matching
coefficient is
\begin{equation}\label{3pjmatching}
 \Gamma^{\alpha \mu\sigma \delta}(M)  = 4 g_s e e_b  
(g^{\perp}_{\alpha \delta} g^{\perp}_{\mu \sigma} + 
g^{\perp}_{\alpha \sigma} g^{\perp}_{\mu \delta} 
- g^{\perp}_{\alpha \mu} n^\delta n^\sigma) \,.
\end{equation}

Next we give Feynman rules for the color-singlet operator at order
$g^2_s$, and order $g_s^3$ To obtain the $g^2_s$ Feynman rule we first
expand Eq.~(\ref{3s1op1}) to order $g_s^2$
\begin{eqnarray}\label{3s1opex1}
{\cal O}(1,{}^3S_1) &= & 
\sum_p \chi^\dagger_{-{\bf p}} \Lambda\cdot\bsigma_\delta \psi_{\bf p}
 \bn\cdot q \bn\cdot q' \Gamma^{\alpha \beta \delta \mu} 
  ( \bn\cdot q, \bn\cdot q')
\nn \\
&&
{\rm Tr} \Big\{ \Big( \frac{\bn\cdot A_{n,q}}{\bn\cdot q} q^\alpha_\perp-
(A^\alpha_{n,q})_\perp \Big) 
\Big( \frac{\bn\cdot A_{n,q'}}{\bn\cdot q'} q^{\prime\beta}_\perp-
(A^\beta_{n,q'})_\perp \Big) \Big\} \,,
\end{eqnarray}
where $\Gamma^{\alpha \beta \delta \mu}$ is given in
Eq.~(\ref{matching1}), which gives the following tree-level Feynman
rule:
\begin{eqnarray}\label{3s1scetfeyn}
{\cal A}(1,{}^3S_1) &= &  \eta^\dagger_{-{\bf p}} 
 \Lambda\cdot\bsigma_\delta \xi_{\bf p}
 \Gamma^{\alpha \beta \delta \mu} ( \bn\cdot q, \bn\cdot q')
\nn \\
&&
 \Big( \frac{\bn\cdot \epsilon_{n,q}}{\bn\cdot q} q^\alpha_\perp-
(\epsilon^\alpha_{n,q})_\perp \Big) 
 \Big( \frac{\bn\cdot\epsilon _{n,q'}}{\bn\cdot q'} q^{\prime\beta}_\perp-
(\epsilon^\beta_{n,q'})_\perp \Big) \delta^{ab} \,.
\end{eqnarray}
Expanding Eq.~(\ref{3s1op1}) to order $g_s^3$ gives
\begin{eqnarray}
{\cal O}_{g} (1,{}^3S_1) &=& i g_s \eta^\dagger_{-{\bf p}} 
 \Lambda\cdot\bsigma_\delta \xi_{\bf p}
 \Gamma^{\alpha \beta \delta \mu} ( \bn\cdot q, \bn\cdot q')
\nn \\ 
& & 
\Big[-q^{\prime}_\perp \cdot (A^a_{n,q})_\perp + q_\perp \cdot q''_\perp
\frac{\bn\cdot A_{n,q}}{\bn\cdot q} \Big] \frac{\bn\cdot A_{n,q'}}{\bn\cdot q'}
\frac{\bn\cdot A_{n,q''}}{\bn\cdot (q^{\prime}+q'')} \,,
\end{eqnarray}
which gives the Feynman rule
\begin{eqnarray}
i{\cal A}_{g}(1, ^3S_1) &=& -\frac{8 e e_b g_s^3}{3M} 
\eta^\dagger_{-{\bf p}} \Lambda\cdot\bsigma^\mu \xi_{\bf p} f^{abc}
\left\{
\frac{\bn^\alpha\bn^\beta}{\bn\cdot(g_1+g_2)}
   \left(\frac{g_{2\perp}^\gamma}{\bn\cdot g_2} - 
         \frac{g_{1\perp}^\gamma}{\bn\cdot g_1}\right)\right.\nonumber\\
&& +
\frac{\bn^\beta\bn^\gamma}{\bn\cdot(g_2+g_3)}
   \left(\frac{g_{3\perp}^\alpha}{\bn\cdot g_3} - 
         \frac{g_{2\perp}^\alpha}{\bn\cdot g_2}\right) + 
\frac{\bn^\gamma\bn^\alpha}{\bn\cdot(g_3+g_1)}
   \left(\frac{g_{1\perp}^\gamma}{\bn\cdot g_1} - 
         \frac{g_{3\perp}^\gamma}{\bn\cdot g_3}\right)\nonumber\\
&& +
\bn^\alpha\bn^\beta\bn^\gamma\left[
\frac{g_{1\perp}\cdot g_{2\perp}}{\bn\cdot g_3}
\left(\frac1{\bn\cdot g_2\bn\cdot(g_1+g_3)} -
      \frac1{\bn\cdot g_1\bn\cdot(g_2+g_3)}\right)\right.\nonumber\\
&&
\phantom{\bn^\alpha\bn^\beta\bn^\gamma +}+
\frac{g_{2\perp}\cdot g_{3\perp}}{\bn\cdot g_1}
\left(\frac1{\bn\cdot g_3\bn\cdot(g_2+g_1)} -
      \frac1{\bn\cdot g_2\bn\cdot(g_3+g_1)}\right)\nonumber\\
&&
\phantom{\bn^\alpha\bn^\beta\bn^\gamma +} +
\left.\left.\frac{g_{3\perp}\cdot g_{1\perp}}{\bn\cdot g_2}
\left(\frac1{\bn\cdot g_1\bn\cdot(g_2+g_3)} -
      \frac1{\bn\cdot g_3\bn\cdot(g_2+g_1)}\right)\right]\right\}.
\end{eqnarray} 

\section{Color-octet Factorization}\label{facappen}

In this Appendix we show factorization for the color-octet
contributions.  At leading order in $\lambda$, matching the current
gives
\begin{equation}\label{currmatch}
J_\mu =  
  i e^{-i(Mv+\bnP \frac{n}{2} )\cdot x} \bigg[ 
\Gamma^{(8,{}^1S_0)}_{\alpha \mu}(M) \,  \tilde{J}^\alpha_{(8,{}^1S_0)} + 
\Gamma^{(8,{}^3P_0)}_{\alpha \mu \sigma \delta} (M) \,
  \tilde{J}^{\alpha \sigma \delta}_{(8,{}^3P_0)} 
\bigg] \,.
\end{equation}
Implicit in this formula is that we are working in a frame where the 
photon momentum defines the $z$ axis, {\it i.e.} $q_\perp = 0$.
The effective theory currents are
\begin{equation}
\tilde{J}^\alpha_{(8,{}^1S_0)}  = \chi^\dagger_{-{\bf p}}  
B^\alpha_\perp \psi_{\bf p},
\hspace{1cm}
\tilde{J}^{\alpha \sigma \delta}_{(8,{}^3P_0)} =  \chi^\dagger_{-{\bf p}}  
B^\alpha_\perp
\Lambda\cdot\widehat{{\bf p}}^\sigma 
  \Lambda\cdot\bsigma^\delta \psi_{\bf p}  \,.
\end{equation}
The Wilson coefficients are given in Eq.~(\ref{1s0matching}) and
Eq.~(\ref{3pjmatching}). Next we insert Eq.~(\ref{currmatch}) into
Eq.~(\ref{fsamp}), and use momentum conservation to set the $\bnP$ in
the phase to $-M$. At this order in $v$ there is no mixing between the two
color-octet currents so cross-terms in the forward scattering
amplitude vanish, and we can write
\begin{equation}
T(z) = \bigg[ H_{(8,{}^1S_0)}(M,\mu) 
   T^{\rm eff}_{(8,{}^1S_0)} (z, \mu)
 + H_{(8,{}^3P_0)}(M,\mu) 
   T^{\rm eff}_{(8,{}^3P_0)} (z, \mu) \bigg] \,,
\end{equation}
where $T^{\rm eff} $ is the forward scattering matrix element in the
effective theory:
\begin{eqnarray}
T^{\rm eff}_{(8,{}^1S_0)}  &=& - i  \int d^4 x 
\, e^{i(M \frac{\bn}{2} -q)\cdot x}
\langle \Upsilon | T \big[ \psi^\dagger_{\bf p'}  
B^\alpha_\perp \chi_{-{\bf p'}}\big] (x) \,
\big[\chi^\dagger_{-{\bf p}} 
B^\beta_\perp \psi_{\bf p}\big] (0) \, 
 | \Upsilon \rangle \, g^\perp_{\alpha \beta},
  \\
T^{\rm eff}_{(8,{}^3P_0)}  &=& - i \int d^4 x 
\, e^{i(M \frac{\bn}{2} -q)\cdot x}
 \frac{1}{3}  \langle \Upsilon | T \big[ \psi^\dagger_{\bf p'} 
B^\alpha_\perp ({\bf p'}\cdot\bsigma) 
  \chi_{-{\bf p'}}\big] (x) \, 
\big[ \chi^\dagger_{-{\bf p}}  
B^\beta_\perp
({\bf p}\cdot\bsigma) \psi_{\bf p}\big] (0)  | 
  \Upsilon \rangle \, g^\perp_{\alpha \beta},
\nn
\end{eqnarray}
where the $\Upsilon$ states in $T^{\rm eff} $ are non-relativistically
normalized. For the sake of simplicity we only consider
the color-octet ${}^1S_0$ contribution. The treatment of the
color-octet ${}^3P_0$ contribution is identical. The usoft gluons in
$T^{\rm eff}$ can be decoupled from the collinear fields using the
field redefinition given in Eq.~(\ref{fieldred}).  Under this
substitution the color-octet ${}^1S_0$ current becomes
\begin{equation}
\tilde{J}^\alpha_{(8,{}^1S_0)}  = \chi^\dagger_{-{\bf p}} Y 
 B^{(0) \alpha}_\perp  Y^\dagger \psi_{\bf p} \,,
\end{equation}
where the collinear fields in $B^{(0)}_\perp$ do not interact with the usoft
fields.  Since $| \Upsilon \rangle$ contains no collinear quanta we can
write the effective forward scattering amplitude as
as
\begin{eqnarray}\label{x11}
T^{\rm eff}_{(8,{}^1S_0)}  &=& \frac{- i}{2}   \int d^4 x 
\, e^{\frac{i}{2}M(1-z)\bn\cdot x}
\langle \Upsilon | T \big[ \psi^\dagger_{\bf p'} 
  Y T^A Y^\dagger \chi_{-{\bf p'}}\big] (x) \,
\big[ \chi^\dagger_{-{\bf p}} 
  Y T^A Y^\dagger \psi_{\bf p} \big] (0)| \Upsilon \rangle
\nn \\
& & \times
\langle 0 | T \,  {\rm Tr}\big\{ T^B B^{(0) \alpha}_{\perp}(x) \big\} 
{\rm Tr}\big\{ T^B B^{(0)\beta}_{\perp}(0) \big\} 
\, g^\perp_{\alpha \beta}  | 0 \rangle \,,
\end{eqnarray}
where $q^\mu = \frac{M}{2} z \bn^\mu$, and  we used 
\begin{eqnarray}
\langle \Upsilon | T \big[ \psi^\dagger_{\bf p'} 
  Y T^A Y^\dagger \chi_{-{\bf p'}}\big] (x) \,
\big[ \chi^\dagger_{-{\bf p}} 
  Y T^B Y^\dagger \psi_{\bf p} \big] (0)| \Upsilon \rangle &=&
\nn \\
&& \hspace{-30ex}
\frac{\delta^{AB}}{8} 
\langle \Upsilon | T \big[ \psi^\dagger_{\bf p'} 
  Y T^C Y^\dagger \chi_{-{\bf p'}}\big] (x) \,
\big[ \chi^\dagger_{-{\bf p}} 
  Y T^C Y^\dagger \psi_{\bf p} \big] (0)| \Upsilon \rangle\,.
\end{eqnarray}
Next we introduce the jet function which is defined as
\begin{equation}\label{jetfun}
\langle 0 | T \,  {\rm Tr}\big\{ T^B B^{(0) \alpha}_{\perp}(x) \big\} 
{\rm Tr}\big\{ T^B B^{(0)\beta}_{\perp}(0) \big\} 
\, g^\perp_{\alpha \beta}  | 0 \rangle \equiv 
i \int \frac{d^4 k}{(2\pi)^4} e^{-ik\cdot x} J_M(k^+),
\end{equation}
where the jet function $J_M(k^+)$ is labeled by the large lightcone
momentum of the jet (which in this case is $M$). It is a function of
only one component of the usoft momentum $k^+$, which follows from the
collinear Lagrangian containing only the $n\cdot \partial$
derivative~\cite{Bauer:2001yt}.  Inserting Eq.~(\ref{jetfun}) into
Eq.~(\ref{x11}), we can integrate over three of the $k$ momentum
components to obtain
\begin{eqnarray}\label{bot1}
T^{\rm eff}_{(8,{}^1S_0)}  &=& 
  \frac{1}{4\pi} \int d^4 x \, \delta(x^+) \delta^{(2)}(x_\perp)
\, \int dk^+ \, e^{\frac{i}{2}(M(1-z)-k^+)\bn\cdot x} J_M(k^+)
\nn \\
& & \qquad \times 
\langle \Upsilon | T \big[ \psi^\dagger_{\bf p'} 
  Y T^A Y^\dagger \chi_{-{\bf p'}}\big] (x) \,
\big[ \chi^\dagger_{-{\bf p}} 
  Y T^A Y^\dagger \psi_{\bf p} \big] (0)| \Upsilon \rangle \,.
\end{eqnarray}
Next we simplify the matrix element of usoft operators. We will use the
following identity~\cite{Bauer:2001yt}
\begin{equation}
YT^A Y^\dagger = {\cal Y}^{BA} T^B \,,
\end{equation}
where ${\cal Y}^{BA}$ is the usoft Wilson line in the adjoint representation
\begin{equation}
{\cal Y}^{BA}(x) = 
\bigg[ {\rm P exp} \bigg( ig \int^x_{- \infty} ds n\cdot A^E_s(ns) 
{\cal T}^E \bigg) \bigg]^{AB} \,,
\end{equation}
with $({\cal T}^E)^{AB} = -i f^{EAB}$.  Furthermore we introduce a
usoft Wilson line of finite length
\begin{equation}
{\cal Y}^{BA}(x) {\cal Y}^{CA}(0) = {\cal Y}^{BC}(0,x) \,,
\end{equation}
where 
\begin{equation}
{\cal Y}^{BC}(0,x) = \bigg[ {\rm P exp} \bigg( ig \int^x_0 ds n\cdot A^E_s(ns) 
{\cal T}^E \bigg) \bigg]^{BC} \,.
\end{equation}
Finally we introduce an octet usoft function defined as
\begin{equation}\label{softfun}
S(\ell^+) = 
 \int \frac{dx^-}{4 \pi} \, e^{\frac{-i}{2} \ell^+ x^-} 
\langle\Upsilon | 
  \big[ \psi^\dagger_{\bf p'}  T^B  \chi_{-{\bf p'}}\big] (x^-) 
  \, {\cal Y}^{BC}(0,x^-)
  \big[ \chi^\dagger_{-{\bf p}}  T^C  \psi_{\bf p} \big] (0)| 
\Upsilon \rangle \,.
\end{equation}
Performing three of the integrals over $x$ in Eq.~(\ref{bot1}), we can
substitute the usoft function as defined above to obtain the desired
factored form
\begin{equation}
T^{\rm eff}_{(8,{}^1S_0)}  = \int d\ell^+ S(\ell^+) J_M[\ell^+ + M(1-z)] \,.
\end{equation}

The final step is to perform an OPE by integrating out the collinear
degrees of freedom. This is done by calculating the jet function order
by order in a perturbative expansion in $\alpha_s(M\sqrt{1-z})$, and
matching the forward scattering amplitude in SCET onto a forward
scattering amplitude that is a convolution of a hard coefficient and a
usoft operator, Eq.~(\ref{nonotmore}).  We already know the usoft
operator we are matching onto, Eq.~(\ref{softfun}). We
carried out the matching and running for the color-octet contribution
in a previous paper~\cite{Bauer:2001rh}, and we will not repeat the
calculation here. Readers are referred to that paper for details.



\begin{thebibliography}{}

\bibitem{bbl}
G.~T.~Bodwin, E.~Braaten and G.~P.~Lepage,
Phys.\ Rev.\ D {\bf 51}, 1125 (1995)
[Erratum-ibid.\ D {\bf 55}, 5853 (1995)]
[hep-ph/9407339].

\bibitem{lmr}
M.~E.~Luke, A.~V.~Manohar and I.~Z.~Rothstein,
Phys.\ Rev.\ D {\bf 61}, 074025 (2000)
[hep-ph/9910209].

\bibitem{Catani:1995iz}
S.~Catani and F.~Hautmann,
Nucl.\ Phys.\ Proc.\ Suppl.\  {\bf 39BC}, 359 (1995)
[hep-ph/9410394].

\bibitem{Maltoni:1999nh}
F.~Maltoni and A.~Petrelli,
Phys.\ Rev.\ D {\bf 59}, 074006 (1999)
[hep-ph/9806455].

\bibitem{Rothstein:1997ac}
I.~Z.~Rothstein and M.~B.~Wise,
Phys.\ Lett.\ B {\bf 402}, 346 (1997)
[hep-ph/9701404].

\bibitem{Bauer:2001ew}
C.~W.~Bauer, S.~Fleming and M.~Luke,
Phys.\ Rev.\ D {\bf 63}, 014006 (2001)
[hep-ph/0005275].

\bibitem{Bauer:2001yr}
C.~W.~Bauer, S.~Fleming, D.~Pirjol and I.~W.~Stewart,
Phys.\ Rev.\ D {\bf 63}, 114020 (2001)
[hep-ph/0011336].

\bibitem{Bauer:2001ct}
C.~W.~Bauer and I.~W.~Stewart,
Phys.\ Lett.\ B {\bf 516}, 134 (2001)
[arXiv:hep-ph/0107001].

\bibitem{Bauer:2001yt}
C.~W.~Bauer, D.~Pirjol and I.~W.~Stewart,
Phys.\ Rev.\ D {\bf 65}, 054022 (2002)
[arXiv:hep-ph/0109045].

\bibitem{Bauer:2001rh}
C.~W.~Bauer, C.~W.~Chiang, S.~Fleming, A.~K.~Leibovich and I.~Low,
Phys.\ Rev.\ D {\bf 64}, 114014 (2001)
[arXiv:hep-ph/0106316].

\bibitem{Fleming:2002rv}
S.~Fleming and A.~K.~Leibovich,
arXiv:hep-ph/0211303.

\bibitem{Photiadis:1985hn}
D.~M.~Photiadis,
Phys.\ Lett.\ B {\bf 164}, 160 (1985).

\bibitem{Hautmann:2001yz}
F.~Hautmann,
hep-ph/0102336.

\bibitem{firstRad}
S.~J.~Brodsky, D.~G.~Coyne, T.~A.~DeGrand and R.~R.~Horgan,
Phys.\ Lett.\ B {\bf 73}, 203 (1978).
K.~Koller and T.~Walsh,
Nucl.\ Phys.\ B {\bf 140}, 449 (1978).

\bibitem{Kramer:1999bf}
M.~Kramer,
Phys.\ Rev.\ D {\bf 60}, 111503 (1999)
[arXiv:hep-ph/9904416].

\bibitem{Petrelli:1997ge}
A.~Petrelli, M.~Cacciari, M.~Greco, F.~Maltoni and M.~L.~Mangano,
Nucl.\ Phys.\ B {\bf 514}, 245 (1998)
[arXiv:hep-ph/9707223].

\bibitem{Buskulic:1995au}
D.~Buskulic {\it et al.}  [ALEPH Collaboration],
Z.\ Phys.\ C {\bf 69}, 365 (1996).

\bibitem{Altarelli:1977zs}
G.~Altarelli and G.~Parisi,
Nucl.\ Phys.\ B {\bf 126}, 298 (1977).

\bibitem{Owens:1986mp}
J.~F.~Owens,
Rev.\ Mod.\ Phys.\  {\bf 59}, 465 (1987).

\bibitem{Bauer:2002nz}
C.~W.~Bauer, S.~Fleming, D.~Pirjol, I.~Z.~Rothstein and I.~W.~Stewart,
Phys.\ Rev.\ D {\bf 66}, 014017 (2002)
[arXiv:hep-ph/0202088].

\bibitem{FL}
S.~Fleming and A.~K.~Leibovich, work in progress.

\bibitem{bc}
E. Braaten and Y.Q. Chen,
Phys.\ Rev.\ {\bf D54}, 3216 (1996).

\bibitem{Gremm:1997dq}
M.~Gremm and A.~Kapustin,
Phys.\ Lett.\ B {\bf 407}, 323 (1997)
[arXiv:hep-ph/9701353].

\bibitem{Bodwin:2001mk}
G.~T.~Bodwin, D.~K.~Sinclair and S.~Kim,
Phys.\ Rev.\ D {\bf 65}, 054504 (2002)
[arXiv:hep-lat/0107011].

\bibitem{Petrelli:1998ge}
A.~Petrelli, M.~Cacciari, M.~Greco, F.~Maltoni and M.~L.~Mangano,
Nucl.\ Phys.\ B {\bf 514}, 245 (1998)
[hep-ph/9707223].

\bibitem{Wolf:2001pm}
S.~Wolf,
Phys.\ Rev.\ D {\bf 63}, 074020 (2001)
[hep-ph/0010217].

\bibitem{Leibovich:2001ra}
A.~K.~Leibovich, I.~Low and I.~Z.~Rothstein,
hep-ph/0105066.

\bibitem{Nemati:1996xy}
B.~Nemati {\it et al.}  [CLEO Collaboration],
Phys.\ Rev.\ D {\bf 55}, 5273 (1997)
[arXiv:hep-ex/9611020].

\bibitem{Hagiwara:pw}
K.~Hagiwara {\it et al.}  [Particle Data Group Collaboration],
Phys.\ Rev.\ D {\bf 66}, 010001 (2002).

\bibitem{Field:cy}
R.~D.~Field,
Phys.\ Lett.\ B {\bf 133}, 248 (1983).

\bibitem{Field:2001iu}
J.~H.~Field,
Phys.\ Rev.\ D {\bf 66}, 013013 (2002)
[arXiv:hep-ph/0101158].

\end{thebibliography}
\end{document}